\documentclass[aps,pra,twocolumn,reprint,superscriptaddress,nofootinbib]{revtex4-1}
\pdfoutput=1
\usepackage{color}
\usepackage[normalem]{ulem}
\usepackage{graphicx}
\usepackage{amsmath}
\usepackage{amsfonts}
\usepackage{amssymb}
\usepackage{microtype}
\usepackage{verbatim}
\usepackage{bbold}
\usepackage{hyperref}
\allowdisplaybreaks
\usepackage{multirow}
\usepackage{bbm}
\usepackage{bm}
\usepackage{braket}
\usepackage{soul}
\usepackage{placeins} 
\usepackage{natbib}
\makeatletter
\newsavebox{\@brx}
\newcommand{\llangle}[1][]{\savebox{\@brx}{\(\m@th{#1\langle}\)}%
  \mathopen{\copy\@brx\kern-0.5\wd\@brx\usebox{\@brx}}}
\newcommand{\rrangle}[1][]{\savebox{\@brx}{\(\m@th{#1\rangle}\)}%
  \mathclose{\copy\@brx\kern-0.5\wd\@brx\usebox{\@brx}}}
\makeatother

\hypersetup{
  colorlinks   = true, 
  urlcolor     = blue, 
  linkcolor    = blue, 
  citecolor   = blue 
}

\usepackage{textcomp}

\usepackage[smallerops]{newtxmath}

\begin{document}

\author{Tamiro Villazon}
\affiliation{Department of Physics, Boston University, 590 Commonwealth Ave., Boston, MA 02215, USA}

\author{Pieter W. Claeys}
\email{pc652@cam.ac.uk}
\affiliation{TCM Group, Cavendish Laboratory, University of Cambridge, Cambridge CB3 0HE, UK}

\author{Anatoli Polkovnikov}
\affiliation{Department of Physics, Boston University, 590 Commonwealth Ave., Boston, MA 02215, USA}

\author{Anushya Chandran}
\affiliation{Department of Physics, Boston University, 590 Commonwealth Ave., Boston, MA 02215, USA}

\title{Shortcuts to Dynamic Polarization}

\begin{abstract}
Dynamic polarization protocols aim to hyperpolarize a spin bath by transferring spin polarization from a well-controlled qubit such as a quantum dot or a color defect. 
Building on techniques from shortcuts to adiabaticity, we design fast and efficient dynamic polarization protocols in central spin models that apply to dipolarly interacting systems. 
The protocols maximize the transfer of polarization via bright states at a nearby integrable point, exploit the integrability-breaking terms to reduce the statistical weight on dark states that do not transfer polarization, and realize experimentally accessible local counterdiabatic driving through Floquet-engineering. 
A master equation treatment suggests that the protocol duration scales linearly with the number of bath spins with a pre-factor that can be orders of magnitude smaller than that of unassisted protocols.
This work opens new pathways to cool spin baths and extend qubit coherence times for applications in quantum information processing and metrology.
\end{abstract}


\maketitle

\section{Introduction}\label{sec:Intro}

A prevalent goal in several fields of physics and chemistry is to efficiently polarize an ensemble of spin particles. In nuclear magnetic resonance spectroscopy (NMR) and magnetic resonance imaging (MRI), polarizing nuclear spins enhances sensitivity and resolution~\cite{thankamony2017dynamic,bowen2008time,gallagher2008magnetic,maansson200613}. In applications to quantum information processing, hyperpolarization schemes can be used to initialize large-scale quantum simulators~\cite{cai_simulator_2013} or to extend qubit coherence times by cooling the surrounding spin bath~\cite{foletti2009universal,london_polar_2013}. Where costly or difficult to polarize the spin ensemble directly, dynamic polarization protocols have been developed to repeatedly transfer polarization from readily polarized control spins~\cite{atsarkin1978dynamic,abragam1982nuclear,maly2008dynamic,thankamony2017dynamic,can2015mechanisms,scheuer2017robust,schwartz2018robust,ajoy2018orientation}.
In simple experimental setups, a spin bath is polarized by controlling a single qubit, such as a nitrogen vacancy (NV) center in diamond~\cite{schirhagl2014nitrogen,belthangady2013dressed,fernandez-acebal_hyper_2017} or a quantum dot~\cite{gullans2013preparation,lai2006knight,urbaszek2013nuclear}, whose polarization can be repeatedly reset, effectively generating a zero temperature reservoir for the bath~\cite{christ2007quantum}. 
A key goal of this article is to introduce a fast and efficient scheme for dynamic polarization in central spin models.  

Polarization transfer relies on the spin-flip interactions between a control spin and the spin ensemble to be polarized. The Hamiltonian can be schematically represented as
\begin{equation}
H = \Omega(t)\, S^z + H_{\textrm{spin-flip}}\, ,
\end{equation}
consisting of an electromagnetic field $\Omega(t)$ acting on the control spin along the z-direction and spin-flip interactions between control spin and spin bath. Given an initially polarized control spin, $\Omega(t)$ can be tuned to transfer polarization \cite{overhauser1953polarization,hartmann1962nuclear,rovnyak2008tutorial}. 
Specifically, dynamic polarization protocols can be separated into two classes: (i) \emph{sudden} protocols in which the control field $\Omega(t)$ is quenched to resonance with the spin-flip interactions to induce polarization transfer and (ii) \emph{adiabatic} protocols in which polarization transfer is induced by slowly driving $\Omega(t)$ across resonances~\cite{can2017ramped}. Adiabatic protocols offer an advantage over sudden protocols as they do not require precise resonance tuning and pulse timing. They also can cover a broader range of bath spin resonances, enabling robust transfer in the presence of field and interaction inhomogeneities
~\cite{hediger1995adiabatic,henstra2014dynamic,tan2020adiabatic}. Their main disadvantage is the requirement of slow speeds, which can be inefficient or unfeasible in experiments limited by spin diffusion in the bath and decoherence of the control spins~\cite{fischer2009spin,ramanathan2008dynamic}.

Apart from the limitations on control speeds, the achievable polarization is also limited by the presence of dark states, making it seldom possible to completely polarize the bath even at slow speeds~\cite{christ2007quantum,imamoglu_optical_2003,taylor_controlling_2003,villazon2020persistent,villazon2020integrability}. 
Dark states are many-body qubit-bath eigenstates in which the qubit is effectively decoupled from the bath.
Since such states have a fixed control spin polarization and cannot be depopulated through changes in the qubit control field, any initial nonzero population of dark states will limit hyperpolarization. 
Experiments in different material systems have found maximum saturation at about $60\%$ full polarization~\cite{bracker2005optical,urbaszek2007efficient,chekhovich2010pumping}.

Several schemes have been proposed to enhance hyperpolarization by depopulating dark states effectively~\cite{urbaszek2013nuclear}, for example by modulating the electron wavefunction of the qubit in quantum dots~\cite{imamoglu_optical_2003,christ2007quantum} or by alternating resonant drives which reduce quantum correlations in the bath~\cite{rao2019spin}. While studies so far mainly focused spin systems where the central spin interacts with its environment through fully isotropic (XXX) interactions, arising in e.g. quantum dots in semiconductors, we consider a model where the interactions are anisotropic (XX), as in resonant dipolar spin systems \cite{hartmann1962nuclear,rovnyak2008tutorial,rao2019spin,lai2006knight,ding2014high,taylor2003long,fernandez-acebal_hyper_2017}. 

Overcoming the requirement of slow speeds in adiabatic protocols is the aim of the field of shortcuts to adiabaticity
~\cite{torrontegui2013shortcuts,guery2019shortcuts}.
Shortcut methods such as counterdiabatic driving (CD) suppress diabatic transitions between the eigenstates of a driven Hamiltonian $H(t)$ by evolving the system with a Hamiltonian $H_{CD}(t)$ containing additional counter terms ~\cite{Demirplak1,Demirplak2,Demirplak3,Berry,delCampo2,Muga,Kolodrubetz}. CD preserves the system's  adiabatic path through state space even during ultra-fast protocols. CD protocols typically require engineering operators which are highly complex and many-body, making them difficult to implement in practice~\cite{Kolodrubetz}. 
Recent progress has focused on reducing the complexity of CD Hamiltonians, for example by mapping them to simpler unitary equivalents~\cite{torrontegui2012shortcuts,ibanez_multiple_2012,bukov2019geometric,deffner2014classical}, or by approximating them with local (few-body) operators~\cite{Sels2017,claeys2019floquet,wurtz_variational_2020}.
The required local operators can be realized through e.g. Floquet-engineering techniques, using high-frequency oscillations to realize the CD Hamiltonian as an effective high-frequency Hamiltonian using only controls present in the original adiabatic protocol~\cite{claeys2019floquet,Petiziol,villazon_heat_2019,boyers2019floquet,zhou_floquet-engineered_2019}. 
Local counterdiabatic driving has recently been realized experimentally in synthetic tight-binding lattices \cite{meier_counterdiabatic_2020}, in IBM's superconducting quantum computer \cite{hegade_shortcuts_2020}, and in a liquid-state NMR system for a nonintegrable spin chain \cite{zhou_experimental_2020}.
Such methods have also gained attention in the context of quantum thermodynamics, where (approximate) CD can be used to speed up underlying adiabatic processes and increase the performance of quantum engines \cite{campo_more_2014,abah_performance_2018,cakmak_spin_2019,funo_speeding_2019,abah_shortcut--adiabaticity_2019,villazon_heat_2019,hartmann_multi-spin_2020,dupays_superadiabatic_2020,abah_shortcut--adiabaticity_2020}.

We develop a dynamic polarization scheme which implements approximate counterdiabatic driving (CD) to quickly and efficiently polarize a spin bath using a tunable qubit while simultaneously depopulating dark states.
In the absence of inhomogeneous bath fields, the model Hamiltoniani is integrable \cite{villazon2020integrability}. We first exploit the integrability of this model to design a CD protocol explicitly targeting all polarization-transferring bright (i.e., not dark) states.
Within all protocols the bright states arise in pairs acting as independent two-level Landau-Zener systems, for which CD protocols can be straightforwardly designed.
While the exact CD protocol targets all bright states and gives rise to a highly involved control Hamiltonian, we show how the CD protocol can be well approximated using local (few-body) operators and experimentally implemented using Floquet engineering (FE). 
In the presence of inhomogeneous bath fields the system is no longer integrable. However, the proposed protocols still lead to a remarkable increase in transfer efficiency. Furthermore, the local counterdiabatic (LCD) protocol dynamically couples dark states to bright states, such that dark states can be depopulated.
Not only are such LCD protocols much easier to implement than the exact CD ones, we find that they outperform CD protocols and lead to a complete hyperpolarization of the spin bath. 

The FE protocols also lead to natural quantum speed limits: there exists a lower bound for the protocol durations below which the FE protocol can no longer accurately mimic the LCD protocol. The emergence of speed limits is ubiquitous in shortcut protocols and control theory~\cite{guery2019shortcuts,bukov2019geometric,Kolodrubetz,deffner_quantum_2017,larocca_quantum_2018,poggi_geometric_2019,garcia-pintos_quantum_2019,abah_shortcut--adiabaticity_2020,puebla_kibble-zurek_2020,hatomura_bounds_2020,campaioli_tightening_2020}. 
Interestingly, our work now suggests that speed limits are also intrinsic in \emph{approximate} local counterdiabatic protocols.

This paper is organized as follows. In Section~\ref{sec:model}, we present the qubit-bath model system and the hyperpolarization scheme used throughout this work. In Section~\ref{sec:protocols}, we construct and detail the CD and LCD protocols and compare their efficiency with unassisted (UA) protocols which do not use shortcut methods.
In Section~\ref{sec:cooling}, we show how our shortcut protocols can be applied to fully polarize a spin bath. A master equation for the hyperpolarization is introduced in Section \ref{sec:LargeL}, which is used to analyze the protocols at large system sizes and show that all protocol durations scale linearly with the number of bath spins.
In Section~\ref{sec:FE}, we show how to realize LCD with FE and discuss the emergence of a quantum speed limit. We conclude in Section~\ref{sec:conclusion} with a discussion of our results in a broader context.

\section{Model and Hyperpolarization Scheme}\label{sec:model}

\subsection{Hamiltonian}

We focus on a concrete central spin model describing a qubit interacting with $L-1$ spin-1/2 bath spins. The Hamiltonian is given by
\begin{equation}\label{eq:H}
H(t) = \Omega_Q(t) \,S_0^z + \sum_{j=1}^{L-1} \Omega_{B,j}\, S_j^z + \frac{1}{2}\sum_{j=1}^{L-1} g_j\, \big(S_0^{+} S_j^{-} + S_0^{-} S_j^{+} \big),
\end{equation}
where $\Omega_Q(t)$ is the magnetic field strength on the qubit, $\Omega_{B,j}$ is the magnetic field strength on the $j^{\textrm{th}}$ bath spin, and $g_j$ is the coupling strength between the qubit and the $j^{\textrm{th}}$ bath spin, with $j=1,2,\dots,L-1$. Eq.~\eqref{eq:H} describes several physical setups in rotating frames, such as color defects or quantum dots coupled to ensembles of nuclear spins via dipolar interactions~\cite{hartmann1962nuclear,fernandez-acebal_hyper_2017,rovnyak2008tutorial,rao2019spin,lai2006knight}.
Spin conserving (`flip-flop') transitions dominate the dipolar interaction provided $g_j \ll\Omega_Q + \Omega_B$, with the latter set by the amplitudes of the continuous driving fields; a standard derivation is given in Appendix~\ref{SI:Model}. The top panel of Fig.~\ref{fig:model} shows a schematic of the model.

Experimentally, the bath field and qubit-bath couplings are spatially inhomogeneous. For simplicity, we model these inhomogeneities as uncorrelated disorder: we draw each $\Omega_{B,j}$ independently from a uniform distribution 
\begin{equation}
\Omega_{B,j}\in[\Omega_B - \gamma_z, \Omega_B + \gamma_z],
\end{equation}
where $\Omega_B$ sets the mean value and $\gamma_{z}$ sets the z-disorder strength.
We also draw each $g_j$ independently from a uniform distribution  
\begin{equation}
  g_j\in  [\overline{g} - \gamma_{xx} , \overline{g} + \gamma_{xx}],
\end{equation}
where $\overline{g}$ sets the mean value and $\gamma_{xx}$ sets the xx-disorder strength. In this work, we probe the weak coupling and disorder regime given by $\gamma_{xx}, \gamma_z < \overline{g} \ll \Omega_B$. 

Since $H$ conserves total magnetization $[H,M] = 0$, where 
\begin{equation}
M \equiv \sum_{j=0}^{L-1} S_j^z,
\end{equation}
its eigenspectrum splits into $L+1$ polarization sectors (see left of lower panel in Fig~\ref{fig:model}). Each sector can alternatively be specified by the number $N = M + L/2$ of spin flips above the fully-polarized state $\ket{\downarrow} \otimes \ket{\downarrow\downarrow\dots\downarrow}$. The aim of hyperpolarization is then to find protocols that systematically reduce $M$, where a fully polarized state corresponds to $N=0$ or $M = -L/2$.

\begin{figure}[ht]
\centering
   \includegraphics[width=1.0\columnwidth]{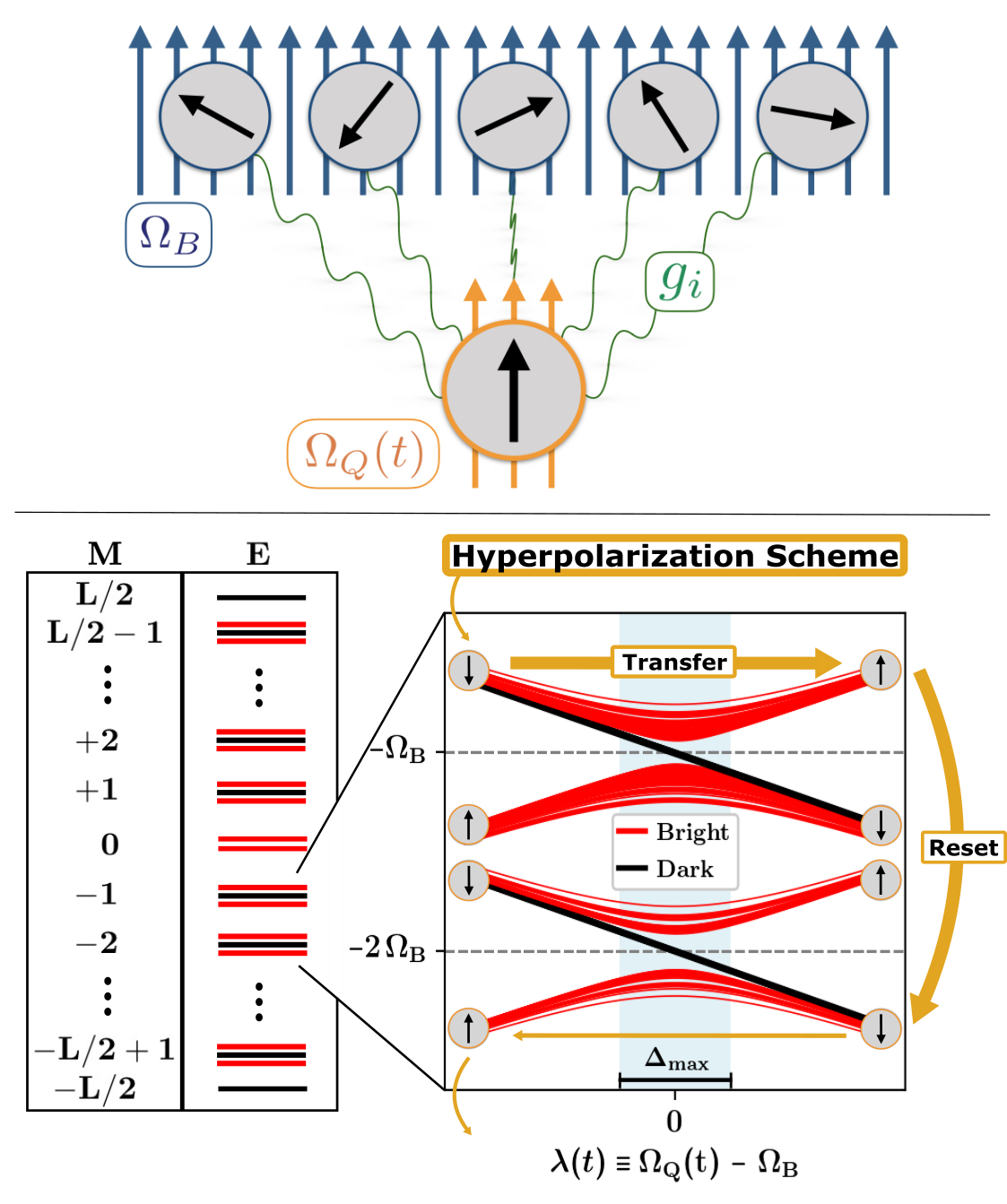}
\caption{\textbf{Model schematic, spectrum, and hyperpolarization scheme.} (Top panel) Schematic of the central spin model in Eq.~\eqref{eq:H}. (Bottom panel) On the left, we illustrate the polarization sectors and spectrum for a system with an even number $L$ of spins and small disorder strengths. On the right, we illustrate the spectrum in two polarization sectors with $M<0$, $\Delta_{\mathrm{max}}$ sets the maximal width of the resonance region, where together with an outline of the transfer-reset hyperpolarization scheme.} \label{fig:model} 
\end{figure}

\subsection{Spectrum}

The eigenstates of $H$ capture essential features common in applications of dynamic polarization: bright states which allow resonant polarization transfer when $\Omega_Q(t)$ is varied, and dark states which limit transfer. In the $\gamma_z = 0$ limit, the model is integrable and the structure of its eigenstates is known~\cite{villazon2020integrability}. 
While our proposed protocols are not restricted to integrable models, the known eigenstate structure at the integrable point allows for a quantitative understanding of the general cooling protocols.
We briefly review these eigenstates and their basic properties in the integrable limit, and subsequently extend the discussion to $\gamma_z > 0$.

\subsubsection{Bright States ($\gamma_z = 0$)}
Bright eigenstates can be written as
\begin{equation}\label{Bright}
\ket{\mathcal{B}^{\alpha} (\lambda) }= c^{\alpha}_{\uparrow}(\lambda)\, \ket{\uparrow} \otimes \ket{\mathcal{B}^{\alpha}_{\uparrow}} + c^{\alpha}_{\downarrow}(\lambda) \ket{\downarrow} \otimes \ket{\mathcal{B}^{\alpha}_{\downarrow}},
\end{equation}
where
\begin{equation}
\lambda(t) \equiv \Omega_Q(t) - \Omega_B
\end{equation} 
measures the detuning between the qubit and bath fields. On resonance, $\Omega_Q = \Omega_B$ and $\lambda = 0$.
The index $\alpha$ distinguishes between the different bright states. 
Crucially, the bath states $|\mathcal{B}^{\alpha}_{\uparrow, \downarrow}\rangle$ do not depend on $\lambda$.

As such, when varying $\lambda$ the bright states only couple in pairs $(\alpha=\pm k)$,  behaving as independent two-level Landau-Zener systems.
The Hamiltonian in each such two-dimensional subspace can be written as
\begin{equation}\label{eq:brightHam}
H_{\alpha}(\lambda) = \lambda \tilde{S}^z_{\alpha} + \Delta_{\alpha}\tilde{S}^x_{\alpha} + \Omega_B\,M,
\end{equation}
where $\Delta_{\alpha}$ sets the energy splitting (gap) of the pair at resonance \cite{villazon2020integrability} and we have introduced generalized spin operators $\tilde{S}_{\alpha}^{x,y,z}$ acting on the two-dimensional space spanned by $\ket{\mathcal{B}^{\alpha}_{+}} = \ket{\uparrow}\otimes \ket{\mathcal{B}^{\alpha}_{\uparrow}}$ and $\ket{\mathcal{B}^{\alpha}_{-}} = \ket{\downarrow}\otimes \ket{\mathcal{B}^{\alpha}_{\downarrow}}$.
\begin{align*}
\tilde{S}^x_{\alpha} &= \frac{1}{2} \left(\ket{\mathcal{B}^{\alpha}_{-}}\bra{\mathcal{B}^{\alpha}_{+}} + \ket{\mathcal{B}^{\alpha}_{+}}\bra{\mathcal{B}^{\alpha}_{-}}\right),\\ 
\tilde{S}^y_{\alpha} &= \frac{i}{2} \left(\ket{\mathcal{B}^{\alpha}_{-}}\bra{\mathcal{B}^{\alpha}_{+}} - \ket{\mathcal{B}^{\alpha}_{+}}\bra{\mathcal{B}^{\alpha}_{-}}\right), \\
\tilde{S}^z_{\alpha} &= \frac{1}{2} \left(\ket{\mathcal{B}^{\alpha}_{+}}\bra{\mathcal{B}^{\alpha}_{+}} - \ket{\mathcal{B}^{\alpha}_{-}}\bra{\mathcal{B}^{\alpha}_{-}}\right).
\end{align*}
$\tilde{S}^z_{\alpha}$ corresponds to $S_0^z$ projected on a bright pair subspace. The apparent simplicity of the problem in this subspace hides the complexity of the qubit-bath interactions present in the original spin basis, where the bath states $\ket{\mathcal{B}^{\alpha}_{\uparrow,\downarrow}}$ and the gap $\Delta_{\alpha}$ are obtained by solving a set of nonlinear Bethe equations \cite{villazon2020integrability}. Within each magnetization sector ${M = N - L/2}$, we label bright state pairs by $\alpha = |k|$, where $k\in \{1,2, \dots, n_B\}$ and 
\begin{equation}\label{eq:n_b}
n_B = {L-1 \choose N-1},
\end{equation}
is the number of pairs in the sector for $M<0$ \footnote{Since the aim of the proposed protocols is to reduce magnetization, we focus on $M <0$.}.

The Hamiltonian \eqref{eq:brightHam} returns the bright state energies
\begin{equation}
E^{\alpha}_{\mathcal{B}}(\lambda) = \Omega_B\,M \pm \frac{1}{2} \sqrt{\,\lambda^2 + \Delta_{\alpha}^2}\,.
\end{equation}
We refer to the set of bright states with positive ${E^{\alpha}_{\mathcal{B}} - \Omega_{B} M > 0}$ as the \emph{top} bright band, and those with negative $E^{\alpha}_{\mathcal{B}} - \Omega_{B} M < 0$ as the \emph{bottom} bright band (see red bands in bottom panel of Fig.~\ref{fig:model}). 

In bright states, polarization can be transferred between the qubit and the bath. At resonance ($\lambda=0$) the bright state pairs are fully hybridized with $c^{\pm k}_{\uparrow} = \pm c^{\pm k}_{\downarrow}$. Initializing the system in a fully polarized central spin state and then quenching to resonance, as is done in sudden protocols, transfers polarization on the timescale $\Delta_{\alpha}^{-1}$.
Adiabatic protocols induce a qubit-bath polarization transfer in bright states by slowly varying $\lambda(t)$ resonance. As $\lambda \to \pm\infty$, bright states approach a product form and the initial eigenstate $\ket{\uparrow} \otimes \ket{\mathcal{B}^{\alpha}_{\uparrow}}$ is adiabatically connected to $\ket{\downarrow} \otimes \ket{\mathcal{B}^{\alpha}_{\downarrow}}$ and vice versa.

Each Landau-Zener problem is fully characterized by its gap. The distribution of bright pair gaps at resonance $\Delta_{\alpha}$ is shown in Fig.~\ref{fig:Gaps} for various polarization sectors. 
As shown in Appendix~\ref{SI:HomoGaps}, the gap distribution can be obtained analytically at zero disorder ($\gamma_{xx}=0$) in the thermodynamic limit where we take $L\to\infty$, holding ${m\equiv |M|/L}$ fixed.
Since these gaps set the necessary time scales for adiabatic protocols, we briefly detail some relevant gap scales.
The typical gap scale is given by
\begin{equation}
\Delta_{\mathrm{typ}} \equiv \sqrt{\sum g_j^2} \sim \sqrt{L-1}\,\,\overline{g},
\end{equation}
shown as a gold vertical dashed line in Fig.~\ref{fig:Gaps}, whereas the maximal gap, also setting the maximal width of the resonance region, is given by 
\begin{equation}
\Delta_{\mathrm{max}}\sim L\,\overline{g},
\end{equation}
scaling extensively in $L$. The smallest bright gap in the homogeneous model can be found as 
\begin{equation}
\Delta_{\mathrm{min}} = \overline{g}\sqrt{2(M+1)} \sim \overline{g}\sqrt{2\,m\,L}.
\end{equation}
At sufficiently small gaps $\Delta \gtrsim \Delta_{\mathrm{min}}$, the distribution of bright pair gaps is given by
\begin{equation}\label{eq:AnalyticGaps}
n(\Delta) \propto \Delta\,\bigg(\frac{1+2m}{1-2m}\bigg)^{-(\Delta/\Delta_{\mathrm{min}})^2} ; \quad \Delta \geq \Delta_{\mathrm{min}}.
\end{equation}
This distribution is shown in Fig.~\ref{fig:Gaps} as a dashed black curve. We find good qualitative agreement between the analytical curve at ${\gamma_{xx}=0}$ and our numerical results for small but finite disorder ${\gamma_{xx}=0.05}$ in the magnetization sector $M=-1$ with the largest Hilbert space dimension.

Fig.~\ref{fig:Gaps} also shows the numerically obtained distribution of bright pair gaps in the $M=-4$ and $M=-7$ sectors. The distribution in the $M=-4$ sector exhibits three broad peaks which are centered around the three bright pair gap energies in the $\gamma_{xx}=0$ limit (Appendix~\ref{SI:HomoGaps}). As the width of each peak is proportional to $\gamma_{xx} L$ while the bright pair gaps at $\gamma_{xx}=0$ are order one, we expect the three-peak structure to be washed out at larger $L$, and the distribution to be captured by Eq.~\eqref{eq:AnalyticGaps} instead. In the $M=-7$ sector, we expect a single pair of bright states with pair gap $\approx \Delta_\mathrm{typ}$ at small $\gamma_{xx}$, as confirmed by Fig.~\ref{fig:Gaps}. We note that Eq.~\eqref{eq:AnalyticGaps} only applies to sectors with finite magnetization density at large $L$.

The main difference comes from the non-zero density of gaps for $\Delta < \Delta_{\mathrm{min}}$. However, as will be shown in following sections, this non-zero density does not qualitatively influence our protocols.

\begin{figure}[htb]
\centering
   \includegraphics[width=1.0\columnwidth]{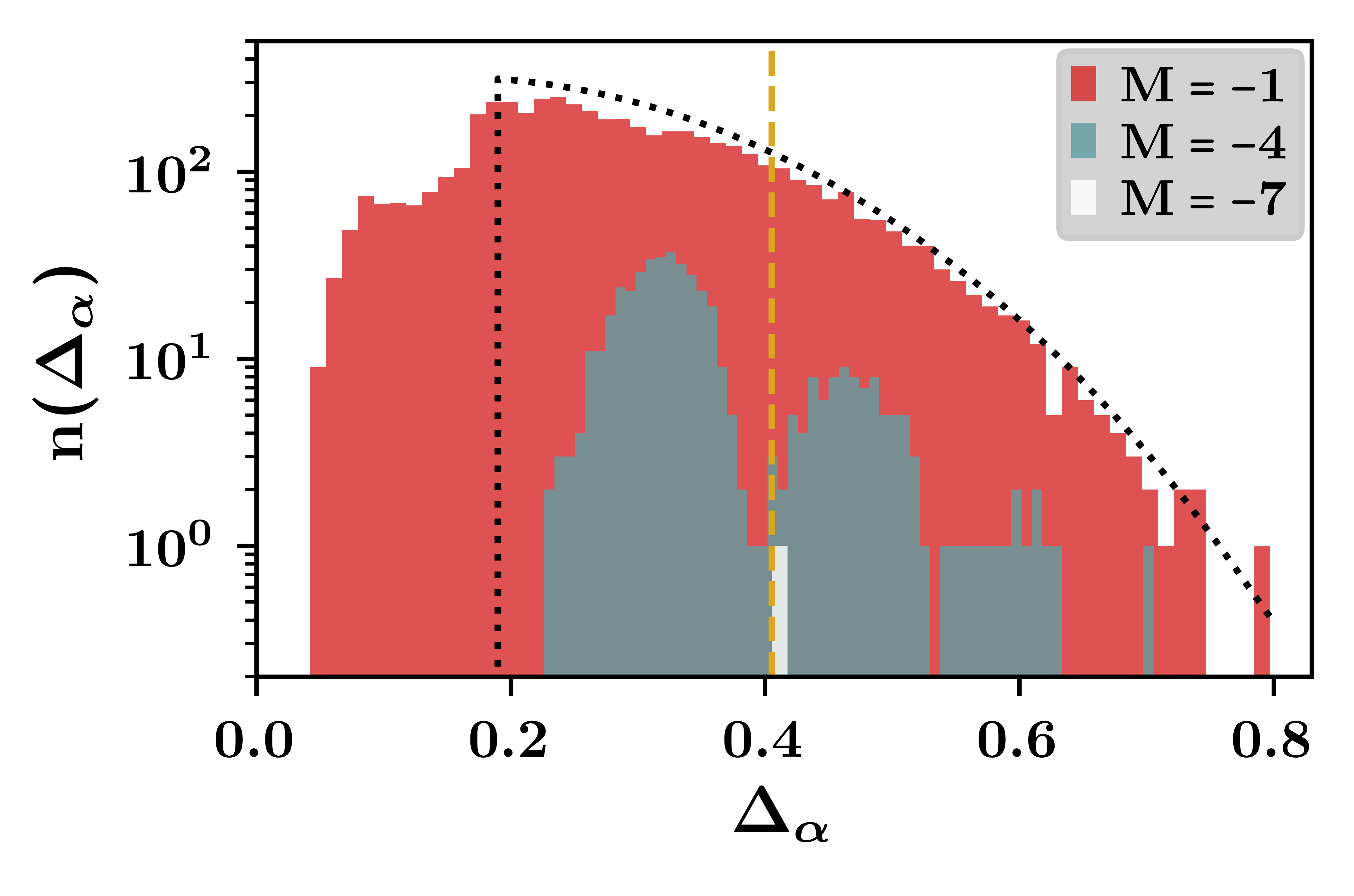}
\caption{\textbf{Distribution of bright pair gaps at resonance.} Histogram of number $\mathrm{n}(\Delta_{\alpha})$ of bright state pairs with gap $\Delta_{\alpha}$. Data is shown for a typical disorder realization in multiple polarization sectors. The gold vertical dashed line denotes the typical gap $\Delta_{\mathrm{typ}}$. The black dashed curve denotes the distribution of gaps from Eq.~\eqref{eq:AnalyticGaps}. Parameters: $L=16$, $\lambda=0$, $\overline{g}=0.1$, $\gamma_{xx}=0.05$, $\gamma_z=0$, and $60$ bins.}  \label{fig:Gaps} 
\end{figure}

In sum, the bright bands consist of an ensemble of independent Landau-Zener systems with a non-trivial distribution of gaps. For each bright pair, an adiabatic passage of $\lambda$ across resonance prevents excitations across its gap and flips polarization of the qubit, transferring polarization to the bath.

\subsubsection{Dark States ($\gamma_z=0$)}

Dark states have the following product form with the central qubit fully polarized:
\begin{equation}\label{Dark}
\ket{ \mathcal{D}^{\alpha}} = \ket{\uparrow} \otimes \ket{\mathcal{D}^{\alpha}_{\uparrow}}\quad \text{or} \quad \ket{\mathcal{D}^{\alpha}} =\ket{\downarrow} \otimes \ket{\mathcal{D}^{\alpha}_{\downarrow}},
\end{equation}
where the index $\alpha$ distinguishes between the different dark states. 
The bath states $\ket{\mathcal{D}^{\alpha}_{\uparrow,\downarrow}}$ depend implicitly on $\{g_j\}$, but crucially not on $\lambda$, and are obtained by solving a set of `dark' Bethe equations~\cite{villazon2020integrability}. 

In a given polarization sector $M=N-L/2$, there are 
\begin{equation}\label{eq:n_d}
n_D = \left|{L-1 \choose N} - {L-1 \choose N-1 } \right|
\end{equation}
dark states. Dark states with central qubit polarized along $+z$ only exist in sectors $M>0$, while dark states with central spin polarization along $-z$ only exist in sectors $M<0$, and no dark states exist in the sector $M=0$. Dark state are eigenstates of $S_0^z$ and are annihilated by the interaction part of the Hamiltonian \cite{villazon2020integrability}, such that the energies given by
\begin{equation}
E^{\alpha}_{\mathcal{D}}(\lambda) = \Omega_B\,M + \mathrm{sgn}[M] \, \frac{\lambda}{2},
\end{equation}
change linearly with the qubit field detuning $\lambda$. Their wave functions however do not change with $\lambda$, preventing polarization transfer to the bath.

\subsubsection{Bright \& Dark States ($\gamma_z>0$)}

In the presence of z-disorder ($\gamma_z>0$), the system is not integrable. However, the same qualitative picture for the eigenstates holds: on adiabatically changing the detuning $\lambda$ and comparing the polarization of the central spin far away from resonance ($\lambda = \pm \infty$), there exists a subset of `bright states' in which the polarization is changed and a subset of `dark states' for which the polarization is unchanged.

Since the central spin is polarized far away from resonance, a counting argument can be used to determine the number of bright and dark states. Consider a sector with magnetization $M<0$ and dimension $n_{M}={L \choose N}$: there are $n_\downarrow = {L-1 \choose N}$ states in which the qubit is fully polarized along the $-z$-direction and $n_\uparrow = {L-1 \choose N-1}$ states in which the qubit is fully polarized along the $+z$-direction. Far from resonance, the energies are given by $\Omega_B M - \lambda/2$ and $\Omega_B + \lambda/2$ respectively. Comparing the total number of states in the top and bottom band far away from resonance, there must be $n_D = n_{\downarrow} - n_{\uparrow}$ dark states in which the polarization of the qubit does not flip for an adiabatic passage across resonance (see also Fig.~\ref{fig:model}). The remaining $n_M - n_D = 2\,n_{\uparrow}$ states are bright states in which the spin of the qubit flips during such an adiabatic process, consistent with Eqs.~\eqref{eq:n_b} and \eqref{eq:n_d}.
This simple counting argument only uses conservation of total z-magnetization and produces the same qualitative eigenstate band structure as one sthe Bethe ansatz in the integrable limit ($\gamma_z=0$)~\cite{villazon2020integrability}. 

There are, however, important differences between the integrable ($\gamma_z=0$) and non-integrable ($\gamma_z>0$) models in the resonance regime.
When $\gamma_z > 0$, the simple product state structure of dark states and the Landau-Zener structure of bright states is no longer exact: the non-integrable eigenstates are mixtures of the unperturbed states and exhibit ergodic behavior (Appendix~\ref{SI:Chaos}).

While adiabatic protocols transferring polarization in the inhomogeneous model are qualitatively similar to those in the homogeneous model, and bright and dark states can generally be defined by their central spin polarization far away from resonance, non-adiabatic effects can enhance polarization transfer in the inhomogeneous model. 
Finite z-disorder is useful for the purposes of dynamic polarization: in a non-adiabatic protocol dark states can be excited to bright states since $|\langle \mathcal{D}^{\alpha}|S_0^z|\mathcal{B}^{\alpha'}\rangle| > 0$.
Dark states in the inhomogeneous model can be depopulated during a non-adiabatic passage across resonance, such that the limit on hyperpolarization can be overcome by preferentially inducing transitions from dark states to bright states.

\subsection{Hyperpolarization Scheme}

We now discuss the basic hyperpolarization scheme as illustrated in Fig.~\ref{fig:model}.

To polarize the spin bath, we apply a cyclical scheme. In each cycle, we (i) \emph{reset} the polarization of the qubit to $\ket{\downarrow}$ at large detuning, and (ii) we \emph{transfer} polarization from the qubit to the bath by sweeping the central field detuning $\lambda(t)$ across resonance over a timescale $\tau_r$. The reset step is a routine experimental step in quantum computing platforms; for example, in a NV set-up, the qubit can be reset using a rapid optical pulse~\cite{sushkov_magnetic_2014,boyers2019floquet}. The ramp varies $\lambda(t)$ from an initial value $\lambda_i$  to a final value $\lambda_f=-\lambda_i$, such that the cycle starts and ends far from resonance ${\lambda_0\equiv|\lambda_i|=|\lambda_f| \gg \Delta_{\mathrm{max}}}$, where the qubit is completely polarized in every eigenstate. 
From one cycle to the next, the direction of the ramp is reversed (after each reset) in a forward-backward fashion.

During a single reset and sweep cycle probability is transferred in every magnetization sector\footnote{The reset and transfer steps have an effect on all polarization sectors simultaneously.} ($M$) from states with up qubit polarization $\ket{\uparrow}$ in sector $M$ to states with down qubit polarization $\ket{\downarrow}$ in the magnetization sector ($M-1$) (as depicted in Fig.~\ref{fig:model}).
The effects on a single bright state can be readily understood: suppose the system is initially in a bright eigenstate $\ket{\downarrow} \otimes \ket{\mathcal{B}^{\alpha}_{\downarrow}}$, factorizable far away from resonance and with fixed magnetization $M$. Then the total magnetization of the bath state is necessarily $M+1/2$. After an ideal adiabatic transfer across resonance, this bright state is given by $\ket{\uparrow} \otimes \ket{\mathcal{B}^{\alpha}_{\uparrow}}$, again far away from resonance. 
From conservation of magnetization, the bath state now has total magnetization $M-1/2$. Following the reset step of the central spin, this state is reset to $\ket{\downarrow} \otimes \ket{\mathcal{B}^{\alpha}_{\uparrow}}$, which is no longer an eigenstate but rather a superposition of eigenstates. Crucially, these states all have magnetization $M-1$: the total bath magnetization has been reduced. Dark states of the form $\ket{\downarrow}\otimes\ket{\mathcal{D}^{\alpha}_{\downarrow}}$ are left invariant by these steps. After several cycles, dark state populations build up and ultimately saturate the bath spin polarization well above its fully polarized value. 

The success of the protocol depends on the suppression of diabatic excitations.
However, transitions between bright states within their own band are irrelevant for the purposes of polarization transfer, and thus we only require that transitions be suppressed between the bands. Specifically, we mimic a slow smooth ramp $\lambda(t)$ with ramp time $\tau_r \gg \tau_0$, where 
\begin{equation}\label{eq:tau0}
\tau_{0} = 2\lambda_0/\Delta_{\mathrm{min}}^2
\end{equation}
sets the timescale for the onset of diabatic transitions between eigenstate bands (Appendix~\ref{SI:LZ}). 
While such a protocol may still be too slow in practical applications, here it serves only as a starting point which guarantees efficient transfer.

\section{Polarization Transfer Protocols}\label{sec:protocols}
We detail how to speed up adiabatic ramps with the assistance of CD and LCD protocols in a single sweep. Such (L)CD protocols can be exactly analyzed in the integrable limit.
We further compare our CD protocols to unassisted (UA) protocols which, unlike CD, attempt to polarize the bath without engineering additional controls. 
A full cooling protocol consisting of repeated sweeps will be analyzed in Section~\ref{sec:cooling}.

\begin{figure}[ht!]
\centering
   \includegraphics[width=1.0\columnwidth]{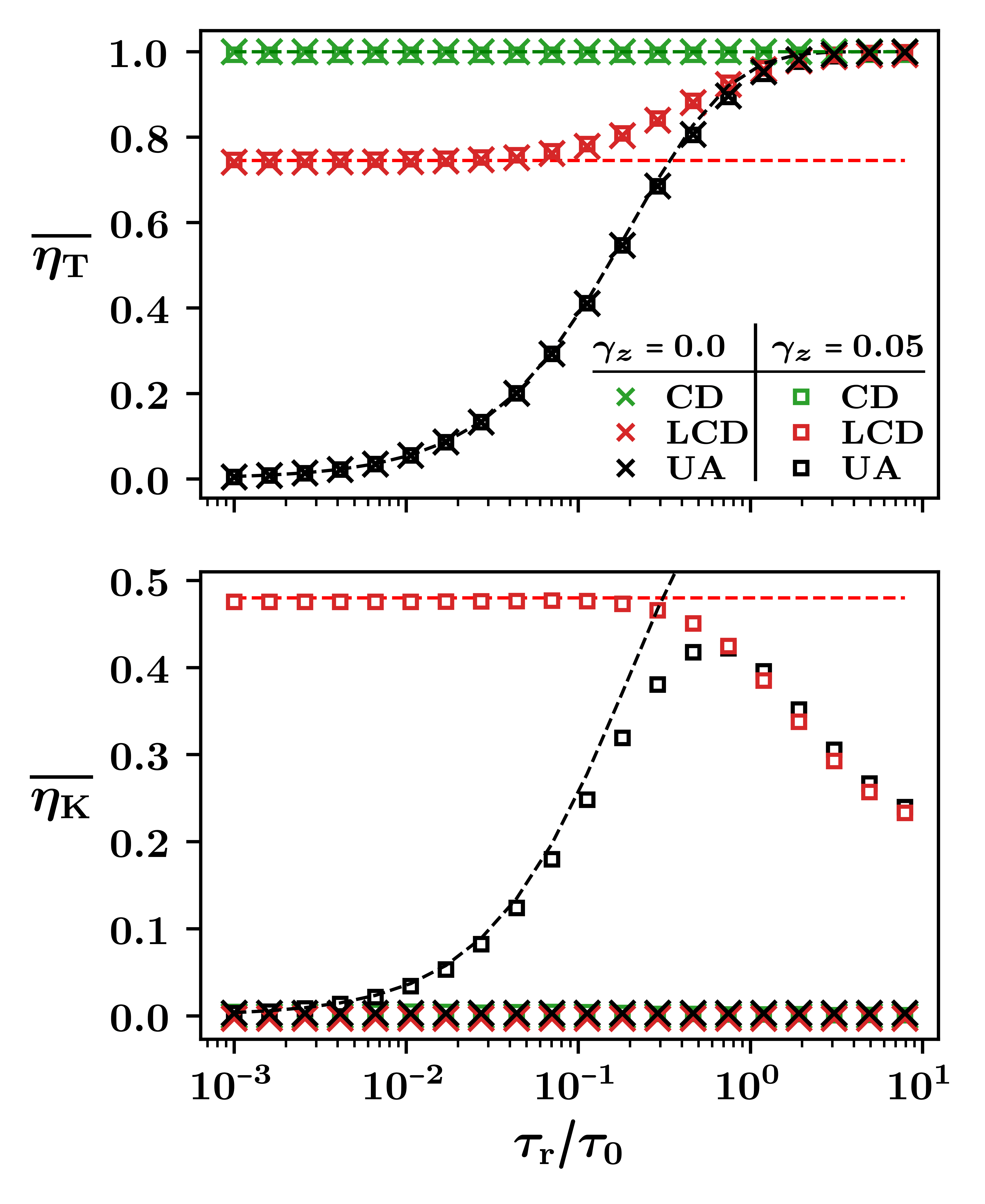}
\caption{ \textbf{Efficiency vs. ramp time.} Disorder-averaged transfer efficiency (top) and kick efficiency (bottom) of UA, CD, and LCD protocols in systems with $\gamma_z=0.00$ (crosses) and $\gamma_z=0.05$ (boxes). Dashed lines show the analytic prediction for the UA transfer efficiency \eqref{eq:AnalyticTransfer}, the analytic predictions for LCD transfer \eqref{eq:AnalyticLCD} and kick efficiencies \eqref{Eff-KT} at large ramp velocities. Parameters: $N_s = 150$, $L=10$, $M = -1$, $\Omega_{B} = 10$, $\lambda_0 = 5$, $\overline{g}=0.1$, $\gamma_{xx} = 0.05$, and $\tau_0 \approx 1000$.} \label{fig:Efficiency} 
\end{figure}

We simulate sweeps $\lambda(t)$ across resonance by numerically solving the time-dependent Schr\"odinger equation \footnote{The specific ramp function used in this work is a smooth polynomial $\lambda(t) = \lambda_0\,(12\,(t/\tau_r)^5 - 30\,(t/\tau_r)^4 + 20\,(t/\tau_r)^3 - 1)$, which monotonically increases from $\lambda(0)=-\lambda_0$ to $\lambda(\tau_r)=\lambda_0$ and has vanishing first and second derivatives at the protocol boundaries. The minimal order of a polynomial in $t/\tau_r$ that satisfies these constraints is five. However, any form $\lambda(t)$ with sufficiently smooth boundary conditions can be used~\cite{Kolodrubetz}.} in a specific polarization sector and measure efficiency. The system is initialized at $\lambda_i = -\lambda_0 \ll - \Delta_{\mathrm{typ}}$ in a mixed state:
\begin{equation}
\rho(t=0) = \ket{\downarrow}\bra{\downarrow} \otimes \rho_{B},
\end{equation}
where the bath is in an infinite-temperature state $\rho_{B}$. This choice of a spatially uncorrelated and unpolarized bath state is motivated by experimental conditions. We also expect any coherences in the initial bath state to be lost during the repeated cycling of the qubit. This gives an initial probability $P_{B\downarrow}(t=0)$ of starting in the top bright band and $P_{D\downarrow}(t=0)$ of starting in the dark band, with
\begin{align}
P_{B\downarrow,\uparrow}(t) &= \sum_{\alpha} \mathrm{Tr}[\rho(t) \, \mathcal{P}_{\downarrow,\uparrow} |\mathcal{B}^{\alpha}\rangle \langle \mathcal{B}^{\alpha} | \mathcal{P}_{\downarrow,\uparrow}], \\
P_{D\downarrow,\uparrow}(t) &= \sum_{\alpha} \mathrm{Tr}[\rho(t) \, \mathcal{P}_{\downarrow,\uparrow} |\mathcal{D}^{\alpha}\rangle \langle \mathcal{D}^{\alpha} | \mathcal{P}_{\downarrow,\uparrow}],
\end{align}
in which $\mathcal{P}_{\downarrow}\equiv \ket{\downarrow}\bra{\downarrow} \otimes I$ is the projection operator to the subspace with down qubit polarization and similarly $\mathcal{P}_{\uparrow}\equiv \ket{\uparrow}\bra{\uparrow} \otimes I$. 
At the end of the ramp ($\lambda_f = +\lambda_0$), the state $\rho(t=\tau_r)$ has a probability $P_{B\uparrow}(t=\tau_r)$ of being in the top bright band, $P_{D\downarrow}(t=\tau_r)$ of being in the dark band, and $P_{B\downarrow}(\tau_r)$ of having transitioned to the bottom bright band. For protocol efficiency, we use two measures: (i) the transfer efficiency,
\begin{equation}\label{eq:TransferEff}
\eta_T \equiv P_{B\uparrow}(\tau_r)/P_{B\downarrow}(0),
\end{equation}
which measures how effectively the qubit polarization in bright states is flipped during a single sweep, and (ii) the kick efficiency,
\begin{equation}\label{eq:KickEff}
\eta_K \equiv 1-P_{D\downarrow}(\tau_r)/P_{D\downarrow}(0),
\end{equation}
which measures how effectively dark states are depopulated (or `kicked') into the bright manifold. Throughout this section, we continually refer to Fig.~\ref{fig:Efficiency}, which plots these efficiencies over a range of ramp times $\tau_r$ for numerically simulated UA and CD protocols. Note that we average over $N_s$ realizations of disorder in $\Omega_{B,j}$ and $g_j$, which we denote by an overline as $\overline{\eta_T}$ or $\overline{\eta_K}$.

\subsection{Unassisted Driving (UA)}

We first discuss unassisted (UA) protocols, corresponding to a sweep of $\lambda$ over a finite time. 
Adiabatic protocols correspond to infinite ramp times $\tau_r \to \infty$, where all bright state polarization is transferred across a single sweep: $\eta_{T}=1$ while $\eta_{K}=0$. 
At finite ramp times diabatic effects become important and generally $\eta_{T} < 1$ and $\eta_{K}>0$.
In the fast limit ($\tau_r \to 0$) the system does not have time to respond to the drive, completely preventing polarization transfer and dark state depletion such that $\eta_T,\eta_K\to0$. 

In a system with a homogeneous bath field ($\gamma_z = 0$), the operator $S_0^z$ only couples bright state pairs\footnote{Note that dark states at $\gamma_z=0$ are eigenstates of $S_0^z$, so they cannot couple to bright states on changing the qubit z-field in time.}, and within each $M$ sector all excitations induced by a finite ramp speed $\dot{\lambda} > 0$ occur only between bright state pairs \cite{villazon2020integrability}.
Each bright state pair can be treated as an independent two-level Landau-Zener problem following Eq.~\eqref{eq:brightHam}, for which the known transition probability for a ramp $\lambda(t)$ across a resonant gap $\Delta_{\alpha}$ is given by~\cite{Shevchenko,AP1}
\begin{equation}
p_{\mathrm{trans}}[\Delta_{\alpha}] = \exp\bigg(-\frac{\pi}{2}\,\frac{\Delta_\alpha^2}{\dot\lambda}\bigg).
\end{equation}
Averaging this transition probability over the gap distribution \eqref{eq:AnalyticGaps} returns an approximate transfer efficiency for a given magnetization sector
\begin{equation}
\overline{\eta}_T = 1 - \frac{\int_{\Delta_{\mathrm{min}}}^{\infty} p_{\mathrm{trans}}[\Delta]\, n(\Delta)\,d\Delta}{\int_{\Delta_{\mathrm{min}}}^{\infty} n(\Delta) \,d\Delta},
\end{equation}
which can be evaluated to return
\begin{align}\label{eq:AnalyticTransfer}
    \overline{\eta}_T = 1 - \frac{\dot{\lambda} \tau_m}{1+ \dot{\lambda} \tau_m} \exp\left(-\frac{m \pi \overline{g}^2 L}{\dot{\lambda}}\right), 
\end{align}
in which $m = M/L$ is the magnetization density, $\dot{\lambda}\propto \lambda_0/\tau_r$, and
\begin{align}
    \tau_m = \frac{1}{m \pi \overline{g}^2L}\ln\left(\frac{1+2m}{1-2m}\right).
\end{align}
The transfer efficiency in Eq.~\eqref{eq:AnalyticTransfer} is plotted in Fig.~\ref{fig:Efficiency} as a dashed black curve and shows excellent agreement with the UA calculations for $\gamma_z=0$ and $\gamma_z=0.05$.

As shown in Fig.~\ref{fig:Efficiency}, the distinction between a system with a homogeneous bath field (crosses) and an inhomogeneous bath field (squares) has little impact on the UA transfer efficiency. 
The transfer efficiency in UA dynamics varies drastically with $\tau_r$. When $\tau_r$ is sufficiently large ($\tau_r \gg \tau_0$; cf. Eq.~\eqref{eq:tau0}), transitions between eigenstate bands become suppressed and the system becomes effectively adiabatic for the purposes of polarization transfer: the qubit flips in bright state bands, but not in dark bands. Fig.~\ref{fig:Efficiency} shows the tendency of simulated UA protocols toward unit transfer efficiency.

The difference between homogeneous and inhomogeneous systems is important when considering the kick efficiency. In a system with inhomogeneous bath fields ($\gamma_z>0$), $S_0^z$ couples bright and dark eigenstates. As such, inhomogeneous fields lead to a nonzero kick efficiency because diabatic transitions can depopulate dark states, whereas homogeneous fields lead to a zero kick efficiency at all ramp rates.

The convergence $\overline{\eta}_T \to 1$ (shown) occurs much faster than the convergence $\overline{\eta}_K \to 0^+$ (not shown).
The former is determined by the gap between bright state pairs, which remain finite throughout the ramp at numerically accessible system sizes, whereas the latter is determined by the dark-bright gaps, which tend to close away from resonance. This leads to dark-bright transitions at large yet finite $\tau_r \lesssim (\Delta E)^{-2}$, where $\Delta E$ is on the order of the level spacing (Appendix~\ref{SI:LZ}).
Any attempt to drive the system faster ($\tau_r \lesssim \tau_0$) leads to diabatic excitations between eigenstates. When $\gamma_z=0$, speeding up UA protocols decreases the transfer efficiency due to transitions between bright bands, but again does not deplete dark states. At finite disorder strength $\gamma_z=0.05$, UA protocols suffer a similar loss of transfer efficiency, but gain the ability to kick dark states, with a peak kick efficiency at intermediate speeds $\tau_r \sim \tau_0$.

\subsection{Exact Counterdiabatic Driving (CD)}

CD protocols suppress transitions between the eigenstates of an instantaneous Hamiltonian by evolving the system with an assisted Hamiltonian that exactly cancels all diabatic excitations~\cite{Kolodrubetz}. The inclusion of counterdiabatic terms in a hyperpolarization protocol can hence be used to increase the transfer efficiency.

Within each two-dimensional Landay-Zener subspace \eqref{eq:brightHam}, the system remains in an instantaneous eigenstate of $H_{\alpha}(\lambda(t))$ at all times when evolved with a time-dependent Hamiltonian \cite{Kolodrubetz}
\begin{equation}\label{eq:CD_LZ}
H_{\mathrm{CD},\alpha}(t) = H_{\alpha}(\lambda(t)) - \dot{\lambda}(t) \frac{\Delta_{\alpha}}{\lambda(t)^2 +\Delta_{\alpha}^2}\, \tilde{S}^y_{\alpha},   
\end{equation}
where the auxiliary (counterdiabatic) term $\propto \tilde{S}^y_{\alpha}$ exactly cancels diabatic transitions between the bright states for arbitrary ramp speeds provided $\dot{\lambda}=0$ at the beginning and end of the ramp.

CD is realized for the full system if the system is evolved with the time-dependent CD Hamiltonian
\begin{equation}\label{CD}
H_{CD}(t) = H(\lambda(t)) + \dot{\lambda}(t)\, \mathcal{A}_{\lambda}(\lambda(t)),
\end{equation}
where the CD term $\mathcal{A}_{\lambda}$, also known as the adiabatic gauge potential, follows as
\begin{align}
\mathcal{A}_{\lambda}(\lambda) &= - \sum_\alpha \frac{\Delta_{\alpha}}{\lambda^2 +\Delta_{\alpha}^2} \tilde{S}^y_{\alpha}.\label{eq:ALZ}
\end{align}
The summation index $\alpha$ runs over all bright pairs in all magnetization sectors. 
The effect of the counterdiabatic term can be understood in the limit $\dot{\lambda}\to \infty$, where the Hamiltonian reduces to $\dot{\lambda}\mathcal{A}_{\lambda}$ and the evolution operator for a single sweep can be written as
\begin{equation}\label{eq:AevolCD}
 \exp\bigg( i \int_{-\infty}^{\infty} \mathcal{A}_{\lambda}\,d\lambda \bigg) = \prod_\alpha \exp\bigg( -i\,\pi\, \tilde{S}^y_{\alpha} \bigg).
\end{equation}
The gauge potential generates a rotation around the y-axis that exchanges $|\mathcal{B}^{\alpha}_{+}\rangle \leftrightarrow |\mathcal{B}^{\alpha}_{-}\rangle$ when $\lambda$ is swept across resonance, exactly as happens in the adiabatic protocol. 

Alternatively, the gauge potential can be written in closed form as (see Appendix~\ref{SI:CD})
\begin{align}\label{eq:Aclosed}
\mathcal{A}_{\lambda} &= - \frac{i}{4}\, (H-\Omega_BM)^{-2} \,[H,S_0^z].
\end{align}
The first (inverse) term in the product is to be interpreted in the sense of a pseudo-inverse, and the second (commutator) term in the product is given by:
\begin{equation}\label{Comm1}
[ H , S_0^z ] = i\,\sum_j g_j (S_0^x \,S_j^y - S_0^y\, S_j^x). 
\end{equation}

The gauge potential is a complex many-body operator, difficult to compute in theory and even harder to implement in practice \cite{Kolodrubetz}. Only in certain special cases, for example when $\partial_{\lambda} H$ is an integrable perturbation of an integrable model $H$, is this operator sufficiently local~\cite{del_campo_assisted_2012,Kolodrubetz,pandey2020adiabatic}. Fortunately, $\partial_{\lambda}H = S_0^z$ is an integrable perturbation of $H$ in the $\gamma_z = 0$ limit of our present model, and the pair structure of the bright state could be used to immediately write down the adiabatic gauge potential. A similar two-level structure for the gauge potential also arises in integrable free-fermionic systems \cite{del_campo_assisted_2012}.

In a system with an inhomogeneous bath field ($\gamma_z>0$), we can no longer express the adiabatic gauge potential explicitly.
Nevertheless, as CD mimics an adiabatic protocol, the transfer efficiency will be maximal.

Fig.~\ref{fig:Efficiency} showcases the effect of exact CD in a transfer protocol across resonance for systems with $\gamma_z=0$ and $\gamma_z=0.05$ (crosses and squares respectively). In both cases, the complete suppression of bright state transitions yields a maximally efficient transfer protocol $\eta_T = 1$, systematically improving on the UA protocol, while the complete suppression of dark state transitions results in zero kick efficiency $\eta_K = 0$.

\subsection{Local Counterdiabatic Driving (LCD)}

In practice, it is hard to realize exact CD, and we must resort to approximation schemes. 
In this section, we follow the method devised in Ref.~\cite{claeys2019floquet} to develop a local approximation $\mathcal{A}_{\mathrm{LCD}}$ to $\mathcal{A}_{\lambda}$. We refer to assisted driving (see Eq.~\eqref{CD}) with $\mathcal{A}_{\mathrm{LCD}}$ as local counterdiabatic driving (LCD). 

As proposed in Ref.~\cite{claeys2019floquet} and detailed in Appendix~\ref{SI:LCD}, a formal expansion for the adiabatic gauge potential can be found in terms of nested commutators:~
\begin{equation} \label{expansion}
\mathcal{A}_{\lambda} = i \sum_{j=1}^{q} \alpha_j \underbrace{[H,[H,\dots[H}_{2j-1},\partial_{\lambda} H]]].
\end{equation}
For $q \to \infty$ Eq.~\eqref{expansion} reproduces the exact gauge potential. A local approximation for $\mathcal{A}_{\lambda}$ is obtained by truncating the commutator expansion of Eq.~\eqref{expansion} to a desired order $q$, and using a variational minimization scheme \cite{Sels2017} to set the coefficients $\alpha_j$ for $j=1,\dots, q$. 

We focus on the leading order term because it (i) is simple enough to be implemented by Floquet driving on the qubit field (see Section~\ref{sec:FE}) and (ii) is already remarkably effective for polarization transfer. One can always refine the approximation to CD by adding higher-order commutators in Eq.~\eqref{expansion}; the rapid convergence of higher-order LCD to CD is shown in Appendix~\ref{SI:LCD}.

To leading order ($q=1$), we obtain:
\begin{equation}\label{ALCD}
\mathcal{A}_{\mathrm{LCD}}(\lambda) = i\,\alpha_1(\lambda)\, [H,S_0^z], \quad \alpha_1(\lambda) = -\frac{1}{\lambda^2+\Delta^2_{\mathrm{typ}}},
\end{equation}
This scheme leads to a coefficient $\alpha_1(\lambda)$ depending on a single energy scale which coincides exactly with the typical gap $\Delta_{\mathrm{typ}}$, in contrast with exact CD, where the prefactor depends either explicitly \eqref{eq:CD_LZ} or implicitly \eqref{eq:Aclosed} on all bright state gaps $\Delta_{\alpha}$. 
In sum,
\begin{align}\label{eq:LCDHam}
H_{\mathrm{LCD}}(t) &= H(\lambda(t)) + i \dot{\lambda}(t)\,\alpha_1(\lambda(t))\, [H,S_0^z] \nonumber \\
&=H(\lambda(t)) + \frac{\dot{\lambda}(t)}{\lambda(t)^2+\Delta_{\mathrm{\mathrm{typ}}}^2}\,\sum_j g_j (S_0^x \,S_j^y - S_0^y\, S_j^x).
\end{align}
Fig.~\ref{fig:Efficiency} shows the resulting LCD curves as unmarked solid red curves ($\gamma_z=0.0$), and circle-marked red curves ($\gamma_z=0.05$). 
LCD is approximate, $\overline{\eta}_T < 1$ (see top panel of Fig.~\ref{fig:Efficiency}). Nevertheless, LCD's transfer efficiency is high ($\overline{\eta}_T \gtrsim 0.75$) over the whole range of ramp times $\tau_r$, even as $\tau_r\to0$ where UA becomes completely transfer inefficient.

A finer comparison between $\mathcal{A}_{\mathrm{LCD}}$ and $\mathcal{A}_{\lambda}$ can be made for $\gamma_z = 0$ by expressing the gauge potential in the Landau-Zener picture \eqref{eq:brightHam},
\begin{equation}\label{LCD-LZ}
\mathcal{A}_{\mathrm{LCD}}(\lambda) = -\sum_{\alpha} \bigg(\frac{\Delta_{\alpha}}{\Delta_{\mathrm{typ}}}\bigg)\, \frac{\Delta_{\mathrm{typ}}}{\lambda^2 +\Delta_{\mathrm{typ}}^2} \tilde{S}^y_{\alpha}.
\end{equation}
Rather than targeting individual gaps as in Eq.~\eqref{eq:ALZ}, the LCD protocols effectively target a single typical energy splitting scale to suppress diabatic transitions between the bright bands. 
In contrast with Eq.~\eqref{eq:ALZ}, the Lorentzian prefactor of $\tilde{S}^y_{\alpha}$ has a fixed width $\Delta_{\mathrm{typ}}$, which does not vary with bright state gap, and a modulated amplitude $\Delta_{\alpha}/\Delta_{\mathrm{typ}}$. 

This discrepancy introduces polarization transfer errors in LCD at intermediate and fast ramps ($\tau_r \lesssim \tau_0$).
Comparing with Eq.~\eqref{eq:AevolCD}, in the limit $\dot{\lambda} \to \infty$ the LCD protocol again generates a rotation within bright state pairs:
\begin{equation}\label{Mismatch}
 \exp\bigg( i \int_{-\infty}^{\infty} \mathcal{A}_{LCD}\,d\lambda \bigg) = \prod_\alpha \exp\bigg( -i\,\pi\,\frac{\Delta_{\alpha}}{\Delta_{\mathrm{typ}}} \tilde{S}_y^{\alpha}\bigg).
\end{equation}
LCD strongly suppresses transitions between those bright pairs with a gap $\Delta_{\alpha} \approx \Delta_{\mathrm{typ}}$, but otherwise yield only partial suppression. 

From Eq.~\eqref{Mismatch} we can define a mismatch error between CD and LCD for each bright pair with gap $\Delta_{\alpha}$ as
\begin{equation}\label{eq:EffErr}
 \mathcal{E}[\Delta_{\alpha}] = \cos^2\bigg( \frac{\pi}{2}\,\frac{\Delta_{\alpha}}{\Delta_{\mathrm{typ}}}\,\bigg). 
\end{equation}
Averaging over the gap distribution \eqref{eq:AnalyticGaps}, the transfer efficiency at large ramp rates is
\begin{equation}
\overline{\eta}_T = 1 - \frac{\int_{\Delta_{\mathrm{min}}}^{\infty} \mathcal{E}[\Delta]\, n(\Delta)\,d\Delta}{\int_{\Delta_{\mathrm{min}}}^{\infty} n(\Delta) \,d\Delta},
\end{equation}
The saddle-point approximation returns
\begin{align}\label{eq:AnalyticLCD}
\overline{\eta}_T = \frac{\int_{1}^{\infty} dt \, t \sin^2 \left(\frac{\pi}{2} \sqrt{2m} t\right) \left(\frac{1-2m}{1+2m}\right)^{t^2}}{\int_{1}^{\infty} dt\, t \left(\frac{1-2m}{1+2m}\right)^{t^2}}.
\end{align}
This expression agrees with the LCD transfer efficiency in Fig.~\ref{fig:Efficiency} (dashed red line) and will be discussed in more detail in the following section.

Fig.~\ref{fig:Efficiency} (bottom panel) also shows the LCD kick efficiency over several $\tau_r$ orders. 
In the $\gamma_z = 0$ limit, LCD has no effect on dark states, just like UA and CD, again leading to a zero kick efficiency (crosses in bottom panel of Fig.~\ref{fig:Efficiency}).
When $\gamma_z>0$, LCD protocols do not prevent dark-bright transitions and exhibit non-zero kick efficiency. 
Since the bright-dark gap is smaller than the typical bright band gap $\Delta_{\mathrm{typ}}$, especially far from resonance where the bright-dark gap tends to close, LCD allows dark-bright transitions as in UA driving. 

The difference in gap scales gives LCD both the advantages of CD for efficient transfer, and the advantages of diabatic UA for depopulating dark states. 
For slow ramps $\tau_r > \tau_0$, LCD and UA have similar transfer efficiencies as the diabatic transition probabilities are small.
In faster ramps $(\tau_r \lesssim \tau_0)$, bright band transitions are suppressed by LCD but not UA.
Meanwhile, LCD saturates to a maximum kick efficiency for $\tau_r < \tau_0$, in contrast with UA protocols which peak around $\tau_r \sim \tau_0$ and then lose kick efficiency as $\tau_r\to0$. 
The distinction between LCD and UA protocols in this fast limit will be further quantified in Eq.~\eqref{Eff-KT}, following Appendix~\ref{SI:Chaos}, where it is argued that in the limit $\tau_r\to0$ the kick efficiency is proportional to the transfer efficiency. % \pwc{and an analytical prediction for the kick efficiency can be obtained}.

Finally, note that this work focuses on the weak xx-disorder limit $\gamma_{xx} < \overline{g}$ where there is a finite gap between bright bands at numerically accessible system sizes. However, a finite bright pair gap is not necessary to design efficient LCD protocols that need only target a typical gap between the bright bands. In Appendix~\ref{SI:Gapless} we show that LCD maintains high transfer and kick efficiencies in the presence of strong xx-disorder even in the presence of small gap as long as the bulk of the bright spectrum still has a gap $\sim\Delta_{\mathrm{typ}}$.

\subsection{Protocol Efficiency and Polarization Sector}
\label{subsec:prot_eff_pol_sec}

\begin{figure}[ht]
\centering
   \includegraphics[width=1.0\columnwidth]{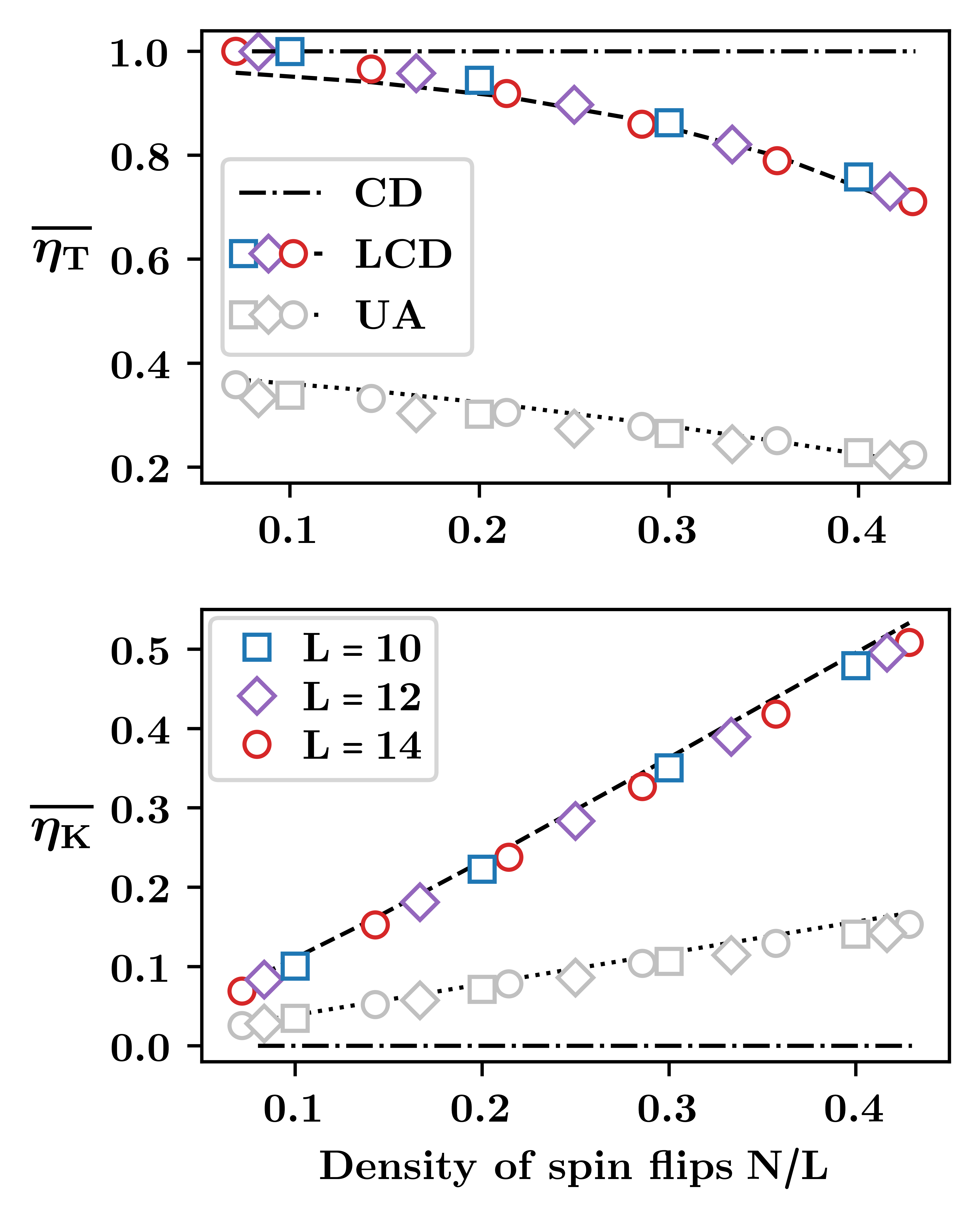}
\caption{ \textbf{Efficiency vs. polarization sector.} The vertical axes show transfer efficiency (top) and kick efficiency (bottom) for fast LCD (colored) and UA (grey) sweeps across resonance in a system with an inhomogeneous bath field. The horizontal axis shows the density $N/L = M/L+1/2$ of spin flips above the fully polarized state in $M<0$ sectors. We plot theoretical predictions to $\overline{\eta}_T$ based on the thermodynamic limit calculations at zero disorder (cf. Eqs.~\eqref{eq:AnalyticTransfer} and \eqref{eq:AnalyticLCD}), and to $\overline{\eta}_K$ based on $\overline{\eta}_T$ (cf. Eq.~\eqref{Eff-KT}). For reference, we also plot corresponding efficiencies for CD protocols. Parameters: $N_s = 150$, $\Omega_{B} = 10$, $\lambda_0= 5$, $\overline{g}=0.1$, $\gamma_{xx} = 0.05$, and $\tau_r = 500/L$.} \label{fig:Eff_Sectors} 
\end{figure}

We compare the efficiencies of CD, LCD, and UA protocols for a single sweep across different polarization sectors $M$. Since all protocols systematically reduce polarization, it is crucial to understand how the transfer and kick efficiencies depend on the polarization sector.
The results are summarized in Fig.~\ref{fig:Eff_Sectors} for a fast ramp ($\tau_r = 0.05\,\tau_0$) in a system with an inhomogeneous bath field ($\gamma_z=0.05$) for multiple system sizes $L$. 

As expected, CD always produces unit transfer efficiency and zero kick efficiency. Moreover, LCD outperforms UA by both efficiency measures in every polarization sector. 
The top panel shows the transfer efficiency $\overline{\eta}_T$, plotted against $N/L = M/L-1/2$.% to collapse the curves for various $L$ onto a single curve. 
For both LCD and UA protocols, the transfer efficiency decreases with $N/L$ because the minimal gap and the number of bright state pairs $n_B = {L-1 \choose N-1}$ increases with $N/L$, in turn increasing the likelihood of diabatic transitions between bright pairs.
For LCD, the transfer efficiency is maximal ($\overline{\eta}_T = 1$) in the sector with $N=1$ because there is only one bright state pair with gap $\Delta_{\mathrm{typ}} = \Delta_{\mathrm{LCD}}$ to target. 

The bottom panel shows the kick efficiency $\overline{\eta}_K$, plotted against $N/L$. % to collapse LCD (UA) curves onto a single curve. 
For both LCD and UA, the kick efficiency increases with polarization.
This increase can be understood by comparing the number $n_D$ of dark states to the number of bright pairs $n_B$ within each sector.
In the sector with $N=1$, there are $(L-2)$ dark states compared to a single pair of bright states, which severely limits the pool of bright states that dark states can transition to. 
As we probe increasingly larger $N$, $n_B$ eventually surpasses $n_D$ such that $n_B/n_D \to \mathcal{O}(L)$ as $N/L \to 0.5$.
The number of available bright states that dark states can transition to increases and thus enhances kick efficiency.

Fig.~\ref{fig:Eff_Sectors} also shows the analytic predictions (black curves) for the transfer efficiency from Eq.~\eqref{eq:AnalyticTransfer} and \eqref{eq:AnalyticLCD}, consistent with the collapse of the curves. For LCD in fast ramps \eqref{eq:AnalyticLCD}, the only dependence of $\overline{\eta}_T$ is on $m = M/L$, consistent with the collapse of $\overline{\eta}_T$ curves at different system sizes as a function of spin flip density ($N/L \sim M/L + 1/2$).
The transfer efficiency in Eq.~\eqref{eq:AnalyticTransfer} depends on both $m$ and $\overline{g}^2 L / \dot{\lambda}$. To achieve a collapse of curves at different system sizes, one must also scale $\dot{\lambda}\sim L$ to eliminate the residual $L$ dependence, yielding a transfer efficiency which depends only on $N/L$. In practice, the collapse can be achieved by scaling up $\lambda_0\sim L$ at fixed ramp-time $\tau_r$ or scaling down $\tau_r\sim1/L$ at fixed ramp range; in our simulations, we have implemented the latter.  
Both predictions show excellent agreement with simulation results in most magnetization sectors, except near $N\sim \mathcal{O}(1)$ where finite size effects are significant. Such finite-size effects also lead to the deviation of the numerically observed bright pair gap distribution in Fig.~\ref{fig:Gaps} from the analytical one at large negative values of $M$.

Remarkably, there is a simple approximate relation between $\overline{\eta}_T$ and $\overline{\eta}_K$ for LCD and UA protocols in the presence of z-disorder at moderate-to-fast ramps $\tau\lesssim \tau_0$. Along with Eqs.~\eqref{eq:AnalyticLCD} and \eqref{eq:AnalyticTransfer}, this relation provides an analytical prediction for the kick efficiency, such that the protocol efficiency can be fully characterized analytically.
Assume that the probability weight that is not successfully transferred to states with up qubit polarization ergodically mixes between the available dark and bright states with spin down. Then,
\begin{equation}
P_{D\downarrow}(\tau_r) \approx P_{D\downarrow}(0) \frac{n_B (1-\overline{\eta}_T) + n_D}{n_B+n_D},
\end{equation}
which implies a kick efficiency
\begin{equation}\label{Eff-KT}
\overline{\eta}_K= 1 - \frac{P_{D\downarrow}(\tau_r)}{P_{D\downarrow}(0)} \approx \frac{n_B}{n_B+n_D}\,\overline{\eta}_T = \frac{N}{(L-N)}\,\overline{\eta}_T.
\end{equation}
The black curves in Fig.~\ref{fig:Eff_Sectors} show $\overline{\eta}_K$ computed using Eqs.~\eqref{Eff-KT}, \eqref{eq:AnalyticLCD} and \eqref{eq:AnalyticTransfer}. These analytic curves are in good agreement with the numerical data for both LCD and UA. Note that the derivation of Eq.~\eqref{Eff-KT} assumes equal mixing between dark and bright states: closer to integrable points, this assumption breaks down, and an analytic relation between the transfer and kick efficiency is no longer possible.

The resulting protocols hence depend on the interplay of different effects across magnetization sectors: increasing $m$, the total number of bright states capable of transfering polarization increases, the average transfer efficiency decreases, and the kick efficiency increases.

\section{Hyperpolarizing the Bath}\label{sec:cooling}

\begin{figure}[ht]
\centering
   \includegraphics[width=1.0\columnwidth]{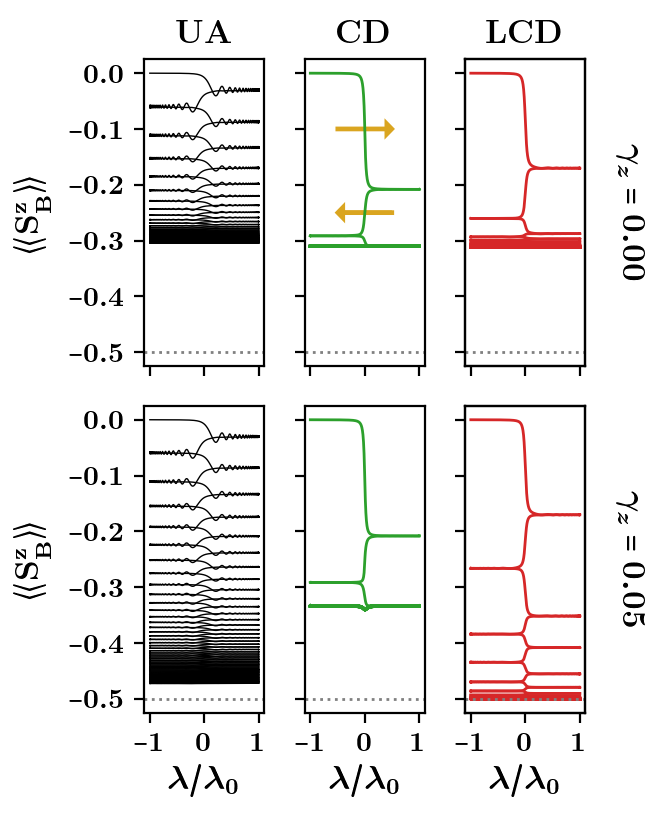}
\caption{ \textbf{Polarizing the spin bath.} Expectation value of the bath polarization per spin $\llangle S^z_B \rrangle$ along a sequence of $N_{\mathrm{c}} =100$ back and forth cycles of the detuning $\lambda$ across resonance. Top and bottom panels show a typical realization for systems with zero and finite z-disorder, respectively. 
The system is initialized in an infinite temperature state.
The forward-backward transfer flow between resets is illustrated by the gold arrows.
Grey dotted lines denote the polarization of the fully polarized state.
Parameters: $L=4$, $\Omega_{B} = 10$, $\lambda_0 = 5$, $\overline{g}=0.1$, $\gamma_{xx} = 0.05$, $\gamma_z = 0$ (top), $\gamma_z = 0.05$ (bottom), and $\tau_r/\tau_0 \approx 0.05$.} \label{fig:Cool} 
\end{figure}

\begin{figure}[ht]
\centering
   \includegraphics[width=1.0\columnwidth]{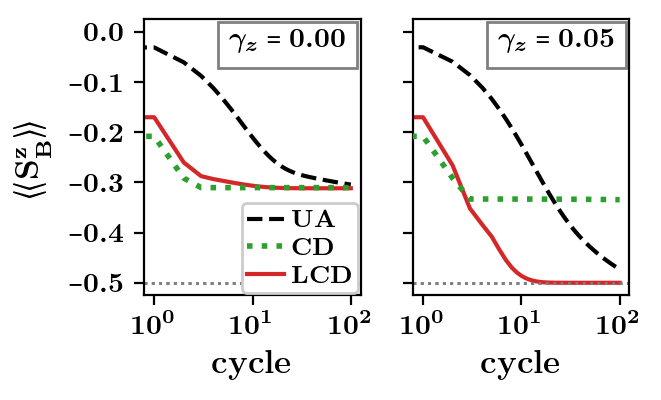}
\caption{ \textbf{Spin bath polarization vs. cycle.} 
Average bath polarization per spin vs. cycle number with the same setup as in Fig.~\ref{fig:Cool}.
Parameters: $N_s=1$, $L=4$, $\Omega_{B} = 10$, $\lambda_0 = 5$, $\overline{g}=0.1$, $\gamma_{xx} = 0.05$, $\gamma_z = 0$ (top), $\gamma_z = 0.05$ (bottom), and $\tau_r/\tau_0 \approx 0.05$.} \label{fig:Cycles} 
\end{figure}

We now turn to the performance of the various protocols over multiple reset-transfer cycles. In particular, we show how the ability of LCD to kick dark states enables complete bath spin polarization. 

Figs.~\ref{fig:Cool} and \ref{fig:Cycles} illustrate the progressive polarization of the bath over multiple ($N_c=100$) reset-transfer cycles in UA, CD, and LCD protocols. We focus on fast sweeps $\tau_r/\tau_0 \approx 0.05$, where the effects of LCD and UA are significantly differentiated. For this simple demonstration, we consider a qubit coupled to 3 bath spins; however, the observed qualitative behavior generalizes to larger baths (see Section \ref{sec:LargeL}). In both figures, we measure the expectation value of the average bath spin polarization per spin: 
\begin{equation}
    \llangle S_{\mathrm{B}}^z \rrangle = \frac{1}{L-1}\sum_{j=1}^{L-1}\langle S^{z}_j \rangle ,
\end{equation}
where $\langle S^{z}_j \rangle \equiv \mathrm{Tr}[\rho(t) S^{z}_j]$ is the expectation value of $S^{z}_j$ in the density matrix $\rho(t)$ of the system at time $t$.

In Fig.~\ref{fig:Cool}, the bath polarization per spin is shown as a function of the detuning $\lambda(t)$ across resonance. After each transfer sweep, the qubit polarization is reset and the direction of the ramp reversed; the resulting forward-backward motion is depicted by the gold arrows. Fig.~\ref{fig:Cycles} shows the corresponding bath polarization per spin after every cycle.

In a typical realization of a system with a homogeneous bath field ($\gamma_z=0$), CD protocols at first quickly reduce the bath polarization due to their maximal transfer efficiency, but soon slow down and saturate as dark states become populated. The saturation point lies well above the fully polarized state (see grey dotted line $\llangle S_{\mathrm{B}}^z \rrangle=-0.5$). In contrast, UA protocols are relatively inefficient and much slower to reach saturation, requiring many more sweeps. LCD protocols perform only slightly worse than CD and much better than UA; they eventually also saturate above the fully polarized state. 

In a typical realization of a system with an inhomogeneous bath field ($\gamma_z=0.05$), CD protocols behave the same as in the homogeneous limit, quickly polarizing the bath to a saturation point. LCD protocols no longer saturate and can polarize the bath close to the fully polarized state due to their non-zero kick efficiency.
Since the hyperpolarization scheme progressively populates smaller $M$ sectors, and the kick efficiency decreases with decreasing $M$ (see Fig.~\ref{fig:Eff_Sectors}), the polarization rate per cycle decreases as we $\llangle S^z_B \rrangle \to -1/2$. UA protocols, like LCD, are able to fully polarize the bath, but their smaller kick efficiency requires many more sweeps.

\section{Scaling to Large Baths}\label{sec:LargeL}

In this section, we explore how the number of cycles needed to hyperpolarize the bath scales with system size $L$.
So far we focused on relatively small system sizes $L \lesssim 10$ to design and test our protocols with accessible exact dynamic simulations.
To circumvent the resource cost of simulating exact dynamics with larger system sizes, we introduce a scalable master equation of the hyperpolarization process in terms of probability flow equations which should be accurate at large transfer speeds.

The state of the system after $c$ transfer-reset polarization cycles is given by:
\begin{align}
    P(c) = 
    \begin{bmatrix} 
    \vec{P}_B\\
    \vec{P}_D
    \end{bmatrix}
    &= \begin{bmatrix}
           P_{B,\downarrow}[0] \\
           P_{B,\downarrow}[1] \\
           \vdots \\
           P_{B,\downarrow}[L] \\
           \hline
           P_{D,\downarrow}[0] \\
           P_{D,\downarrow}[1] \\
           \vdots \\
           P_{D,\downarrow}[L] \\
         \end{bmatrix} \label{Eq:StateBathMaster}
  \end{align}
where $P_{B,\downarrow}[N]$ is the probability of finding the system in a bright state with down qubit polarization in the sector with $N$ spin flips above the fully polarized state, and $P_{D,\downarrow}[N]$ is the probability of finding the system in a dark state with down qubit polarization in the same sector. We do not track the probabilities of bright and dark states with up qubit polarization, as they are always converted to states with down qubit polarization after reset. Moreover, as there exist no dark states with down qubit polarization for $M\geq 0$, $P_{D,\downarrow}[N]=0$ for $N\geq L/2$. Finally, we assume that the bath is fully characterized by the probabilities in Eq.~\eqref{Eq:StateBathMaster} and ignore any correlations in the density matrix between individual dark and bright states since the bath generally decoheres between different polarization cycles. This assumption is justified a posteriori by comparing the efficiencies predicted by the master equation to exact simulations.

\begin{figure}[t]
\centering
 \includegraphics[width=0.85\columnwidth]{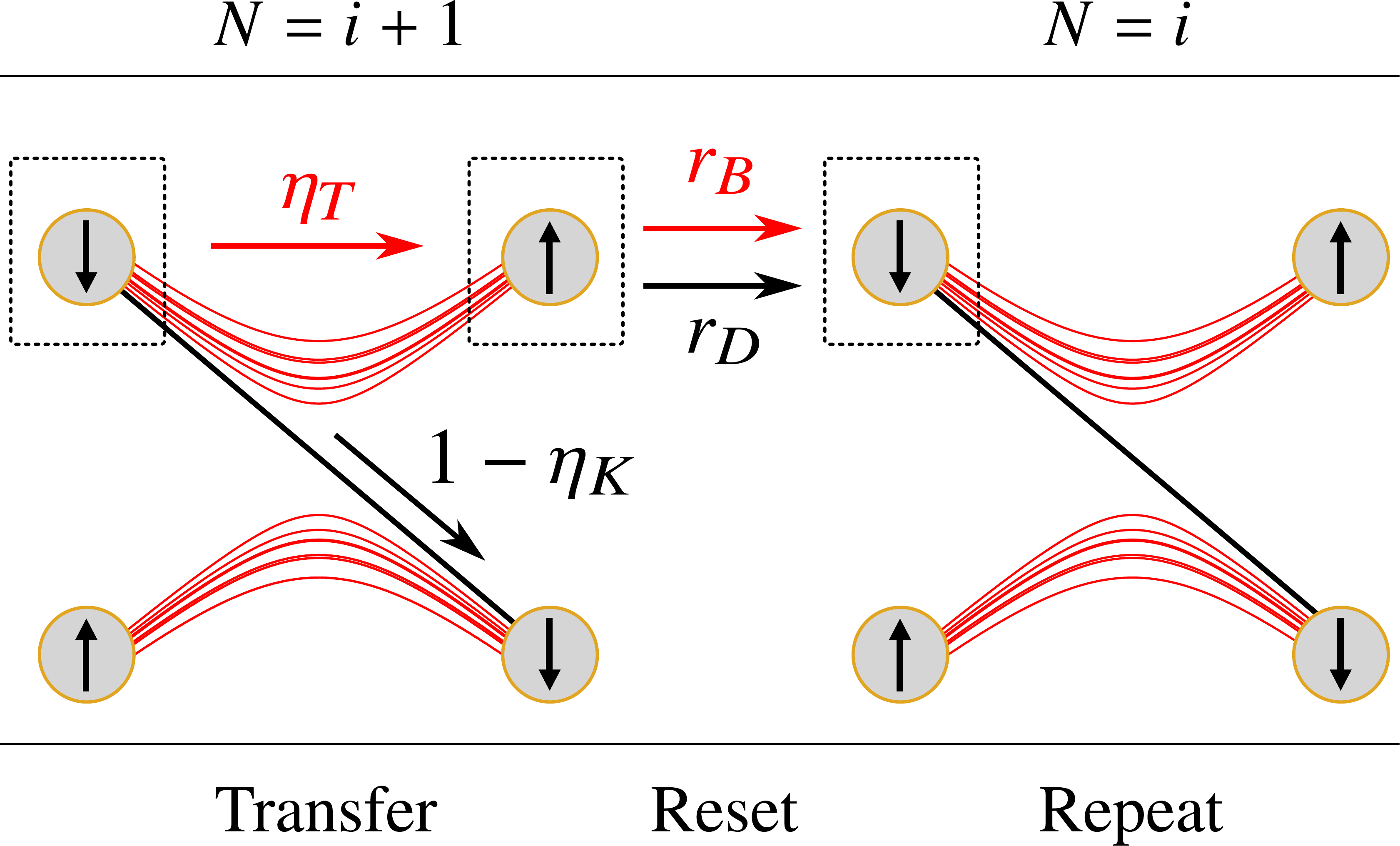}
\caption{\textbf{Effective model}. Schematic representing the action of the transfer matrix through the efficiency functions $\eta$ and reset rates $r$ in bright (red) and dark (black) manifolds in two neighboring polarization sectors $N=i+1$ and $N=i$ during a single polarization (transfer+reset) cycle.}\label{fig:rates} 
\end{figure}

The dynamics of the system is obtained by applying a transfer matrix $T$:
\begin{equation}\label{ToyDyn}
    P(c+1) = T \, P(c),
\end{equation}
where the transfer matrix can be schematically written as
\begin{equation}
T = 
\begin{bmatrix}
    T_{BB} & T_{BD} \\
    T_{DB} & T_{DD}
\end{bmatrix}.
\end{equation}
The transfer efficiency $\eta_T$ sets the probability that the qubit polarization is flipped in bright states during a sweep across resonance. On the other hand, the kick efficiency $\eta_K$ sets the probability that dark states are kicked into bright states. We assume that dark states with down qubit polarization are only kicked into bright states with down qubit polarization.
When the qubit polarization is reset after each sweep, bright states with qubit state $\ket{\uparrow}$ transition to either bright states (with $\ket{\downarrow}$) or dark states (with $\ket{\downarrow}$) in a lower magnetization sector, with relative probability $r_B$ and $r_D$, respectively.
Therefore, the non-zero matrix elements of the transfer matrix are given by:
\begin{align}
    &T_{BB}[i,i] = 1-\eta_T[i] \label{TBBdiag} \\ 
    &T_{BB}[i-1,i] = r_B[i]\,\eta_T[i] \label{TBBoff}\\
    &T_{DB}[i-1,i] = r_D[i]\,\eta_T[i]
    \label{DBoff}\\
    &T_{DD}[i,i] = 1-\eta_K[i]
    \label{TDDdiag}\\
    &T_{BD}[i,i] = \eta_K[i] 
    \label{TBDdiag}
\end{align}
for every sector index $i = 0,\dots, L$. Fig.~\ref{fig:rates} illustrates the transfer and reset rates for a single cycle. Note ${r_B[i]+r_D[i]=1}$, so only one reset rate needs to be specified. %\pwc{ This master equation presents a Markovian approximation of the bath, since we expect the bath state to quickly decohere, and which can be justified a posteriori by comparing the accuracy of the master equation to exact simulations.}

As shown in Section~\ref{sec:protocols}, the different protocols CD, UA, and LCD have different efficiency functions $\eta_T[i]$ and $\eta_K[i]$. 
Here, we consider moderate-to-fast ramp speeds ($\tau_r\lesssim \tau_0$) where LCD and UA have distinct effects. 
Since we are interested in obtaining the scaling of the full protocol, we linearize the previous expressions and model the kick efficiency for LCD and UA as
\begin{equation}
    \eta_K[i] = \eta_0\,\frac{i}{L}; \quad (i\leq L/2),
\end{equation}
and model the corresponding transfer efficiency $\eta_T[i]$ $(i\leq L/2)$ using Eq.~\eqref{Eff-KT}.

The reset rates $r_D[i]$ and $r_B[i]$ depend on the probability distribution of the state within each bright/dark band and details of the structure of eigenstates. Furthermore, these rates can drastically change from one disorder realization to another and are hence difficult to predict. 
To get a reasonable estimate for our master equation, we take $r_D[i+1]$ and $r_B[i+1]$ proportional to to the number of accessible dark and bright states in the $i^{th}$ sector, respectively.
For $M \geq 0$, all the weight is transferred to bright states, 
\begin{equation}
r_B[i+1] = 1, \qquad M \geq 0,
\end{equation}
since dark states have qubit spin up and are not accessible during reset. For $M < 0$, dark states have qubit spin down and become accessible, such that 
\begin{equation}
r_B[i+1] = \frac{n_B[i]}{n_B[i]+n_D[i]}, \qquad M < 0,
\end{equation}
with $n_D$ and $n_B$ given by Eqs.~\eqref{eq:n_b} and \eqref{eq:n_d}, and we refer to this reset rate as well-mixed.
Although we cannot generally satisfy the equiprobable condition within every bright band, we expect to approximately have well-mixed reset rates on average in large disordered systems.     

\begin{figure}[ht]
\centering
   \includegraphics[width=1.0\columnwidth]{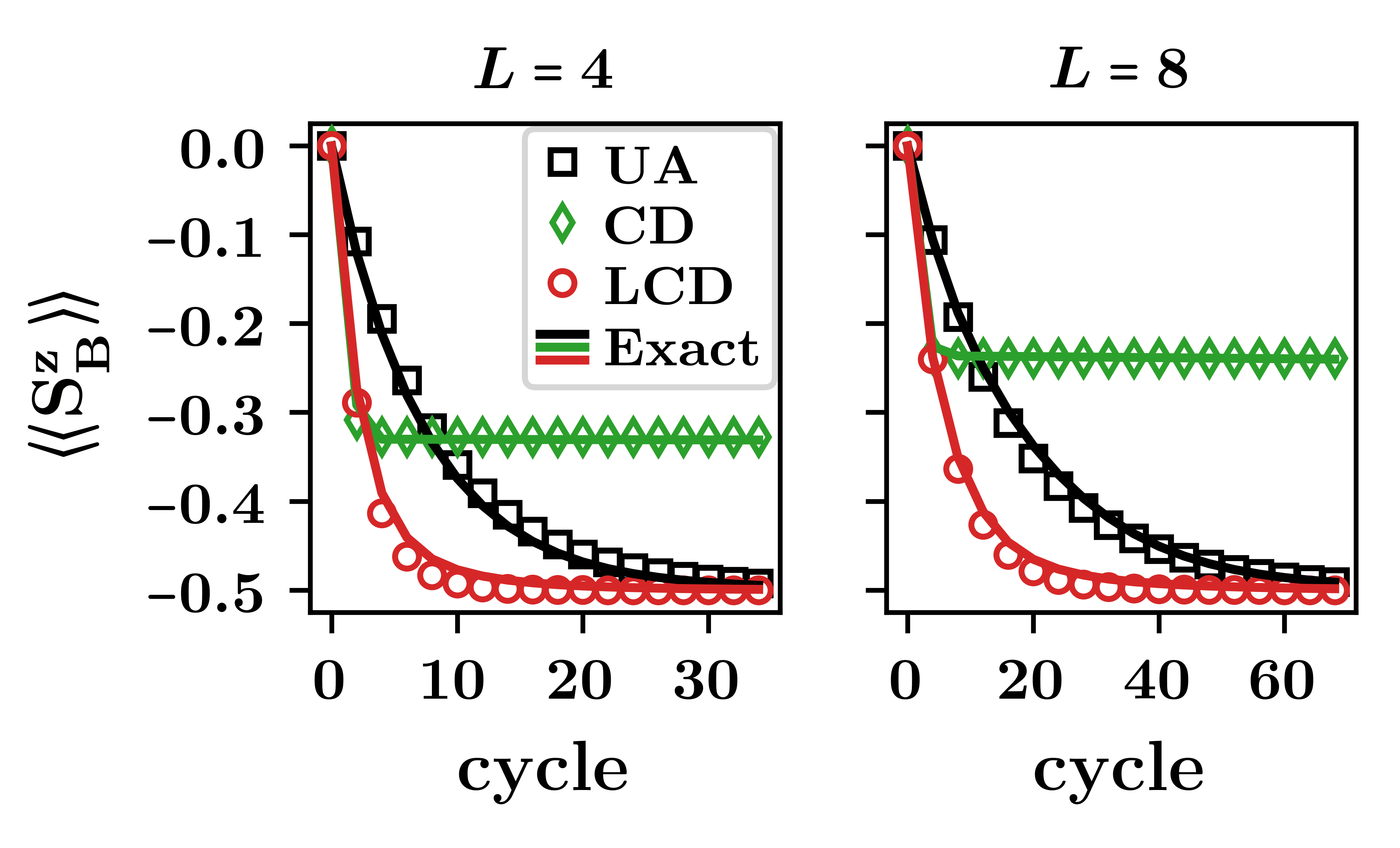}
\caption{ \textbf{Spin bath polarization vs. cycle.} Average bath spin polarization after every transfer-reset cycle over many cycles. The left and right panels show simulation results for systems of size $L=4$ and $L=8$, respectively. Colored markers indicate numerical results using our scalable master equation. Solid colored lines correspond to exact dynamics simulations. 
Parameters: $N_s=1$, $\Omega_{B} = 10$, $\lambda_0 = 5$, $\overline{g}=0.1$, $\gamma_{xx} = 0.05$, $\gamma_z = 0.05$, and $\tau_r = 500/L$.} \label{fig:ToyCycles} 
\end{figure}

We test the master equation in Fig.~\ref{fig:ToyCycles}, which compares UA, CD, and LCD dynamics with Eq.~\eqref{ToyDyn} to the corresponding exact dynamics for different system sizes ($L=4,8$). The figure plots the average bath polarization per spin over many cycles for a single disorder realization. Our master equation agrees well with the results from exact dynamics for all protocols, justifying the assumptions in Eq.~\eqref{Eq:StateBathMaster}. The exact dynamics is computed at small ramp times $\tau_r \approx 0.05\,\tau_0$. To properly capture protocol efficiencies at this ramp speed, we set $\eta_{0} \approx 1.0$ for LCD and $\eta_0 \approx 0.4$ for UA in accordance with the results in Fig.~\ref{fig:Eff_Sectors}. 

\begin{figure}[hb]
\centering
   \includegraphics[width=1.0\columnwidth]{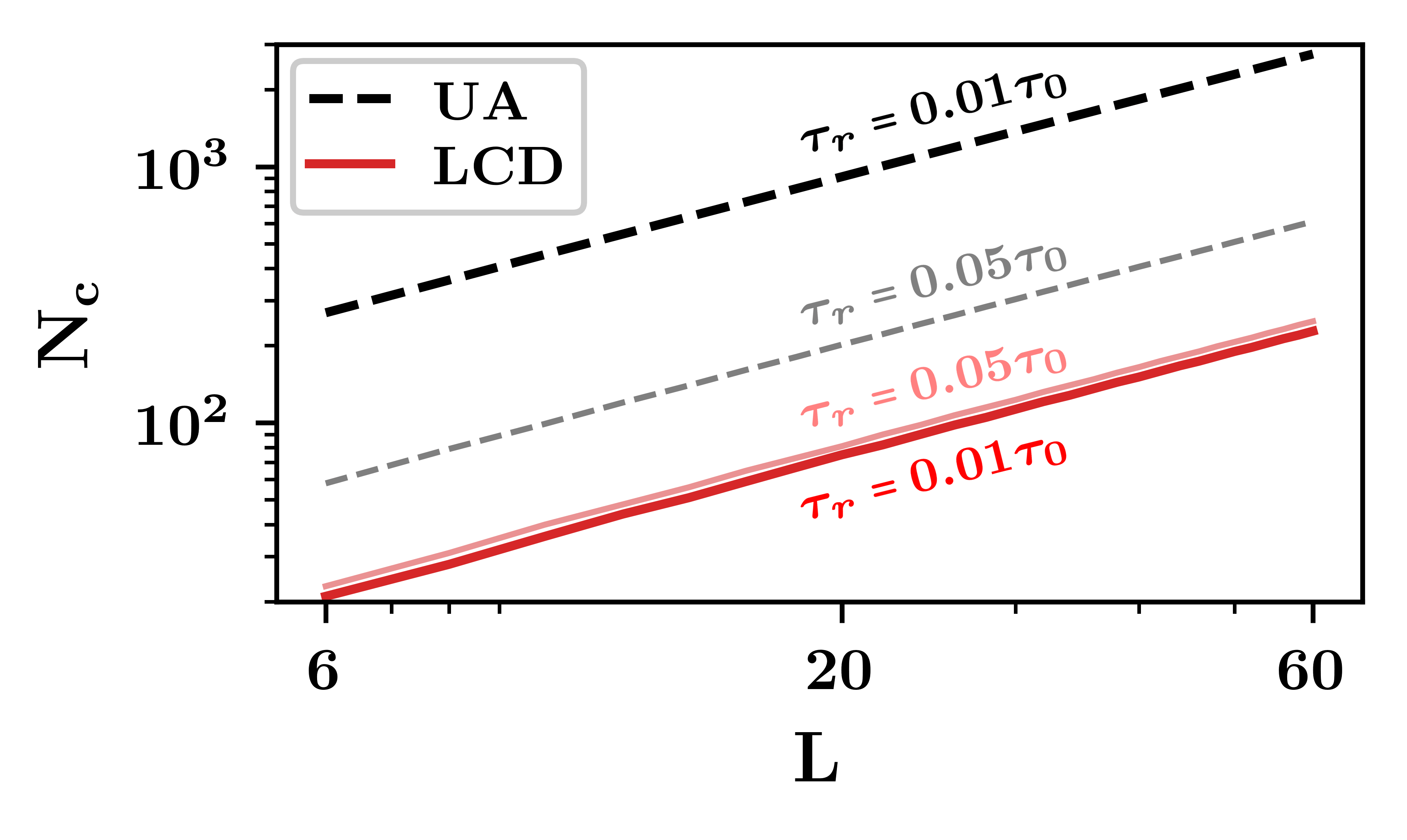}
\caption{ \textbf{Number of cycles to $99\%$ polarization vs. system size. } Master equation simulation results are shown for UA (dashed black) and LCD (solid red) protocol. Opaque curves show results for $\eta_0=1.0$ set for LCD and $\eta_0=0.1$ set for UA, which model ramps with $\tau_r =0.01\,\tau_0$. At this ramp speed, we find $N_c \approx 4 L$ for LCD and $ N_c \approx 40 L$ for UA. Faint curves show results for $\eta_0=1.0$ set for LCD and $\eta_0=0.4$ set for UA, which model ramps with $\tau_r =0.05\,\tau_0$. At this ramp speed, we find $N_c \approx 4 L $ for LCD and $N_c \approx 10 L$ for UA.} \label{fig:Scaling}
\end{figure}

The master equation allows acces to much larger system sizes compared to exact diagonalization. 
Fig.~\ref{fig:Scaling} shows the number of cycles $N_c$ required to reach $99\%$ of the polarization of the fully polarized state against system size $L$. 
The opaque and faint curves denote master equations capturing $\tau_r=0.01\,\tau_0$ and $\tau_r=0.05\,\tau_0$, respectively. 

We find that the number of polarization cycles needed to fully polarize the bath in both LCD and UA protocols scales linearly with system size $L$.
The main difference between LCD and UA is in the prefactor, depending on protocol duration, and which can be orders of magnitude larger in the UA protocol compared to the LCD protocol at sufficiently fast ramps.
In slower ramps $\tau_r>\tau_0$ (not shown), LCD and UA have similar prefactor, but the prefactor for UA however increases as $\tau_r$ is decreased.
Our results are consistent with the expectation that as $\tau_r \to 0$, UA takes progressively more cycles to fully polarize the bath.
Thus, moderate-to-fast LCD is not only time-efficient but also optimizes the number of cycles required to reach the fully polarized state.

We conclude this section with a couple of remarks.
(i) The master equation is applicable at sufficiently fast ramp times $\tau_r < \tau_0 \sim \lambda_0\,\Delta_{\mathrm{min}}^{-2}$, where $\Delta_{\mathrm{min}}\sim \sqrt{L}\,\overline{g}$ in the lowest polarization sectors. To ensure this condition holds as $L\to\infty$, we scale $\lambda_0\sim L$. Otherwise at fixed $\lambda_0$ and sufficiently large $L\sim \lambda_0/(\tau_r\,\overline{g}
^2)$, the master equation would need to be refined to properly account for more complicated speed dependencies in the transfer and kick efficiencies.
(ii) Similarly, our master equation is based on efficiency measurements at sufficiently large z-disorder, where $\gamma_z \gtrsim \Delta_{\mathrm{min}}^{2}/\lambda_0$. In the lowest energy sectors, this requires $\gamma_z\gtrsim L\,\overline{g}^2/\lambda_0$. Again, the $L$ dependence can be cancelled by scaling $\lambda_0\sim L$.

\section{Floquet engineering (FE) of LCD}\label{sec:FE}

The physical implementation of the LCD protocol requires realizing a non-trivial operator $[H, S_0^z]$. We show how it is possible to obtain this LCD Hamiltonian as an effective high-frequency Hamiltonian through Floquet engineering.

Floquet engineering focuses on the design and physical effects of periodic drives~\cite{Bukov2015}. A periodically driven system exhibits dynamics which can be described stroboscopically using an effective slow/static Floquet Hamiltonian $H_F$. Frequently, a control is periodically modulated at a frequency scale $\omega$ larger than any other dynamical frequency in the system, and $H_F$ can be (Magnus) expanded in powers of $\omega^{-1}$~\cite{Bukov2015}. In addition to capturing high-frequency physics, the Magnus expansion has a commutator structure closely related to the structure of the gauge potential in Eq.~\eqref{expansion}, and can be used to realize local counterdiabatic driving at every order~\cite{claeys2019floquet}.

\subsection{Two-level system}\label{subsec:FE:2LS}
In order to give some intuition for the many-body Floquet protocol, we illustrate the general ideas on the two-level system of Eqs.~\eqref{eq:brightHam} and \eqref{eq:CD_LZ}. Temporarily dropping the bright state label and making the time-dependence implicit, the CD Hamiltonian to be realized can be written as
\begin{equation}\label{eq:LZ_CD_1}
H_{\mathrm{CD}} = \lambda \tilde{S}^z + \Delta \tilde{S}^x + \dot{\lambda}\alpha_1 \Delta \tilde{S}^y.
\end{equation}
In an experimental set-up only $\lambda$ is an accessible control parameter, whereas $\Delta$ is constant and the $\tilde{S}^y$ term is absent (as it corresponds to a complex many-body operator acting on the bright pair states). 
In order to realize $H_{\mathrm{CD}}$ as an effective Hamiltonian, we consider a LZ Hamiltonian and add high-frequency oscillations modulated by a slowly-varying amplitude. Specifically, we consider a time-dependent Hamiltonian of the form
\begin{align}\label{eq:HFE_2LS}
H_{\mathrm{FE}}(t) =&\, \gamma(t) \tilde{S}_z + \Delta \tilde{S}_x \nonumber\\
&+\left[{\beta(t)} \omega \sin(\omega t) +{\dot{\beta}(t)}(1-\cos(\omega t)) \right] \tilde{S}_z,
\end{align}
with $\beta(t)$ and $\gamma(t)$ slowly-varying functions to be determined.

The choice of this time-dependence is motivated by the resulting effective Hamiltonian: in the limit of a large driving frequency $\omega$, the stroboscopic dynamics for this time-dependent Hamiltonian is generated by the Floquet Hamiltonian (derived in Appendix~\ref{SI:FE})
\begin{equation}
H_{\mathrm{F}}= \gamma \tilde{S}_z +  J_0(\beta) \Delta  \left[\cos(\beta)\, \tilde{S}_x - \sin(\beta)\, \tilde{S}_y\right],
\end{equation}
where the slow time-dependence has been made implicit and $J_0$ is a Bessel function of the first kind.

The effective Hamiltonian is of the form \eqref{eq:LZ_CD_1}, containing a $\tilde{S}_y$ term not present in the instantaneous Hamiltonian. However, in the CD Hamiltonian the prefactor of $\tilde{S}_x$ is constant and the prefactor of $\tilde{S}_y$ is time dependent. Since the (slow) time dependence of these terms in the Floquet Hamiltonian is determined by the same factor $\beta(t)$, it is not possible to directly realize the CD Hamiltonian in this way. Rather, we can realize a Hamiltonian \emph{proportional to} the CD Hamiltonian.

Demanding $H_{F} = G(t) H_{CD}$, the prefactor for $\tilde{S}_x$ immediately returns the time-dependent prefactor of the full Hamiltonian as
\begin{equation}\label{eq:defGt}
G(t) = J_0(\beta(t)) \cos(\beta(t)).
\end{equation}
Time evolution follows the time-dependent Schr\"odinger equation
\begin{equation}\label{eq:tdse_s}
i \partial_t \ket{\psi(t)} = G(t) H_{\mathrm{CD}} \ket{\psi(t)}.
\end{equation}
Defining a `rescaled time' $s(t)$ such that $\partial_s = G(t) \partial_t$, Eq.~\eqref{eq:tdse_s} can be used to realize counterdiabatic control in the rescaled time provided $i \partial_s \ket{\psi(t(s))} = H_{\mathrm{CD}}(s(t)) \ket{\psi(s(t))}$. The counterdiabatic term is obtained by setting 
\begin{equation}\label{eq:defBeta}
\tan(\beta(t)) = - \alpha_1(s(t)) \dot{\lambda}(s(t)),
\end{equation}
determining $\beta(t)$ as function of $\alpha_1(t)$, leaving
\begin{equation}\label{eq:defGamma}
\gamma(t) = G(t) \lambda(s(t)),
\end{equation}
to finally return $H_F = G(t) H_{\mathrm{CD}}(s(t))$. Note that the experimental time necessarily runs in the positive direction, requiring $G(t) >0$ and $\beta \in [-\pi/2,\pi/2]$.

\subsection{FE protocol}

The ideas in Section~\ref{subsec:FE:2LS} can be immediately extended to the many-body Hamiltonian and LCD of Eq.~\eqref{eq:LCDHam}. 
Given a target LCD ramp with $\lambda(t)=\Omega_Q(t) -\Omega_B$, we drive the system with the Floquet engineered (FE) Hamiltonian: 
\begin{align}\label{FE}
H_{\mathrm{FE}} = H(\Lambda(t)),
\end{align}
with a modified field detuning $\Lambda(t) = \Omega_Q(t) - \Omega_B$ given by
\begin{align}\label{eq:LAMBDA} 
\Lambda(t)=&\, J_0(\beta(t)) \cos(\beta(t))\,\lambda(s(t)) \nonumber\\
&+\beta(t) \,\omega \,\sin(\omega\,t)+\dot{\beta}(t)\,(1-\cos(\omega\,t)).
\end{align}
Following Eq.~\eqref{eq:defBeta}, we set
\begin{equation}\label{eq:beta}
\beta(t) \equiv \arctan\bigg(-\frac{d\lambda(s(t))}{ds}\,\alpha_1(s(t))\bigg),
\end{equation}
and the rescaled time $s = s(t)$ satisfying $ds = G(t) dt$ is defined as
\begin{equation}\label{s}
s = \int_0^{t} J_0 (\beta(t'))\,\cos(\beta(t')) \,dt' > 0.
\end{equation}
This FE Hamiltonian is designed precisely so that the leading order approximation to its Floquet Hamiltonian $H_F$ in the high-frequency limit is the LCD Hamiltonian in the rescaled time:
\begin{equation}\label{HF}
    H_{\mathrm{F}} = H(\lambda(s)) + i\,\frac{d\lambda(s)}{ds}\,\alpha_1(s)[H(\lambda(s)),\partial_{\lambda}H] +\mathcal{O}\bigg(\frac{1}{\omega}\bigg).
\end{equation}
More specifically, the effective Floquet Hamiltonian is found as (see Appendix~\ref{SI:FE})
\begin{align} \nonumber
\tilde{H}_{\mathrm{F}} =&\,G(t)\bigg[\,\lambda(s(t))\,S_0^z + \sum_j \delta\Omega'_j \,S_j^z \nonumber \\ 
&+ \sum_j\,g_j\,( S_0^x \,S_j^{x} + S_0^{y} \,S_j^{y} )-\tan(\beta(t)) \,i\,[H,S_0^z] \bigg],\label{HF-derivation}
\end{align}
where ${\delta\Omega'_j \equiv (\Omega_{B,j}-\Omega_B)/G(t)}$ is the renormalized z-disorder. 

The FE protocol is stroboscopically equivalent to LCD with $\mathcal{A}_{\mathrm{LCD}}$ in Eq.~\eqref{ALCD}. 
Moreover, in a smooth ramp $\lambda$ with $\dot{\lambda}_i = \dot{\lambda}_f = 0$, $H_{FE}$ and $H_{F}$ yield the exact same initial and final states, which guarantees that FE and LCD produce the same polarization transfer during our hyperpolarization scheme.

We remark that $H_F$ equals $H_{\mathrm{LCD}}$ only in the absence of z-disorder ($\gamma_z = 0$). At finite z-disorder ($\gamma_z > 0$), the two differ due to the renormalization $\delta\Omega_j'$ of the bath fields. 
Away from this point, the renormalization tends to enhance z-disorder since $G(t)\in[0,1]$. No significant quantitative differences in performance were found between FE (in lab time) and LCD (in rescaled time) with renormalized disorder, as shown next.   

\begin{figure}[htb]
\centering
   \includegraphics[width=1.0\columnwidth]{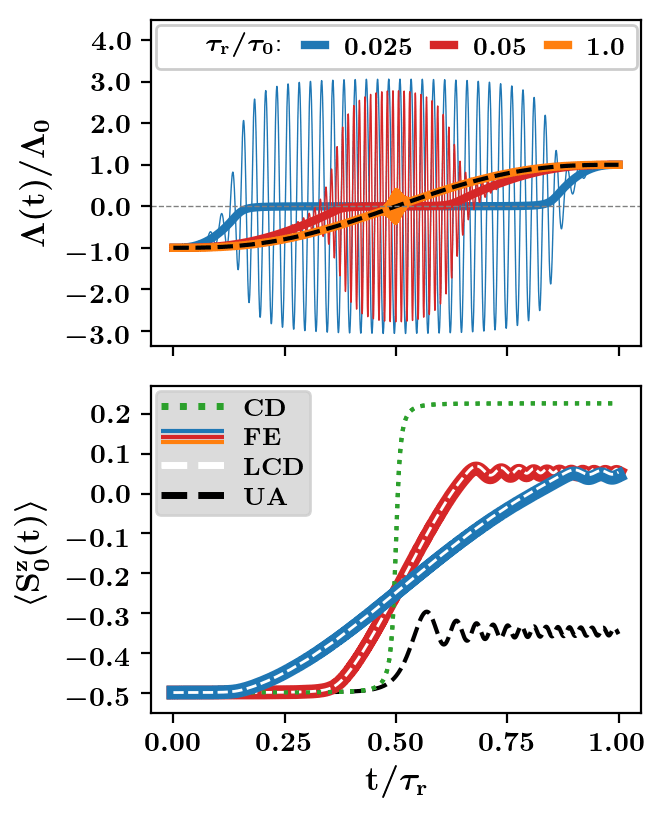}
\caption{ \textbf{Floquet engineered ramp.} Top panel shows the FE ramp $\Lambda(t)$ (solid colored curves) as a function of time $t$ for ramp times $\tau_r = 0.025,0.5,1.0$. The target ramp $\lambda(t)$ (dashed black curve) is shown for reference. The bottom panel shows the effect of FE ramps on the mean qubit polarization $\langle S_0^z(t) \rangle$ over the course of the ramp at $\tau_r = 0.25,0.5$. The corresponding LCD curves are shown to coincide with the FE curves. Curves for UA and CD at $\tau_r = 0.05\,\tau_0$ are shown for reference.  Parameters: $L=8$, $\Omega_{B} = 10$, $\Lambda_0 =\lambda_0= 5$, $\overline{g}=0.1$, $\gamma_{xx} = 0.05$, $\gamma_z = 0.05$, $\tau_0 \approx 1000$, and $\omega=100$ (Note: For display, we have graphically reduced $\omega$ by a factor $10$ to decrease curve density).} \label{fig:Ramp} 
\end{figure}

The upper panel of Fig.~\ref{fig:Ramp} showcases the FE ramp $\Lambda(t)$ in Eq.~\eqref{eq:LAMBDA} for various ramp times $\tau_r$. The vertical axis is re-scaled by the magnitude of the initial/final detunings $\Lambda_0 \equiv \lambda_0$, which are designed to coincide with the target ramp at the ramp endpoints. The target ramp $\lambda(t)$ is shown for reference (dashed black curve). Near the adiabatic breakdown time $\tau_r/\tau_0 = 1$, the FE ramp $\Lambda(t)$ has a base profile (averaging out the oscillations) similar to $\lambda(t)$, with small oscillation amplitudes that get slightly more pronounced in the middle of the ramp around resonance. For progressively faster ramps $\tau_r/\tau_0 = 0.05, 0.025$, the FE ramp $\Lambda(t)$ shows more pronounced deviations from $\lambda(t)$. First observe that the base profile of FE changes, keeping the system near resonance for a progressively larger amount of time. Moreover, the amplitude of the high-frequency oscillations around resonance progressively increases due to $\beta(t)$ in $\Lambda(t)$. Physically, these properties ensure the qubit and bath interact strongly and long enough to effect polarization transfer in accordance with LCD. 

The lower panel of Fig.~\ref{fig:Ramp} serves two purposes: (i) to show the effect of FE on the mean qubit z-polarization $\langle S_0^z \rangle$ over the course of a sweep, and (ii) to highlight the equivalence of FE and LCD protocols. The qubit is initialized with spin down $\langle S_0^z \rangle = -0.5$ in a mixed state. Over the course of the ramp $\Lambda(t)$, FE (solid colored curves) transfers a large fraction of the qubit polarization to the bath in perfect agreement with LCD (dashed white lines). 
For reference we also show an UA protocol at $\tau_r/\tau_0 = 0.05$ (dashed black curve); as expected it is much less efficient compared to FE/LCD and CD at this ramp speed. 

In sum, we can systematically realize LCD with FE, where the LCD protocol can be implemented indirectly in experiments by driving the local qubit field $\Omega_Q(t)$ periodically at high-frequencies $\omega \gg \Omega_Q,\Omega_{B},\tau_r^{-1}$. Importantly, the FE protocol requires no controls which are not already present in $H$ in Eq.~\eqref{eq:H}, similar in spirit to Ref.~\cite{boyers2019floquet}. It can be achieved by setting a fixed global field $\Omega_B$ and dynamically varying $\Omega_Q(t)$, without modifying the qubit-bath interactions. This result differs from other schemes which require controlling interactions to realize LCD with Floquet engineering~\cite{villazon_heat_2019,Petiziol,claeys2019floquet}.

\subsection{Quantum Speed Limit}

The distinction between the lab time $t$ and the rescaled time $s$ gives rise to a quantum speed limit. Namely, there exists a critical ramp time $\tau_r = \tau_{SL} > 0$ in the lab frame for which the protocol duration $\tau_S$ in rescaled time becomes zero and $\beta \to \pi/2$. Given sufficiently large driving frequencies, it is always possible to realize LCD using FE if the LCD ramp time is larger than this critical ramp time. However, at shorter ramp times, the proposed protocol would lead to negative protocol durations in stretched times, and the FE protocol can no longer realize LCD.
The speed limit can derived (see Appendix~\ref{SI:SpeedLimit}) by inverting Eq.~\eqref{s}:
\begin{equation}\label{eq:speedlim}
\tau_{SL} = \lim_{s\to0}\int_0^{s} [G(t(s')]^{-1} ds' \sim \Delta_{\mathrm{typ}}^{-1} .
\end{equation}
The timescale $\tau_{SL}$ is set by the typical bright pair gap $\Delta_{\mathrm{typ}}$, which is on the order of the time needed to transfer polarization from the qubit to the bath while sitting at resonance. In fact, the profile $G(t)\lambda(s(t))$ of the FE protocol in Eq.~\eqref{FE} reduces to a sudden quench protocol to resonance as $\tau \to \tau_{SL}$. 

Eqs.~\eqref{eq:LAMBDA} and \eqref{eq:speedlim} provide an additional connection between \emph{sudden} and \emph{adiabatic} polarization protocols: Floquet-engineering implements the adiabatic polarization protocol in a transformed frame, which resembles a sudden protocol in the lab frame when the speed limit is approached.
For ramps faster than the speed limit $\tau_r/\tau_{SL} < 1$, FE can be extended by taking $\Lambda(t)= \frac{\pi}{2}\,\omega\,\sin(\omega\,t)$; then FE effectively oscillates around resonance for a shorter time than $\tau_{SL}$ and is no longer as effective as LCD.

\begin{figure}[t]
\centering
   \includegraphics[width=1.0\columnwidth]{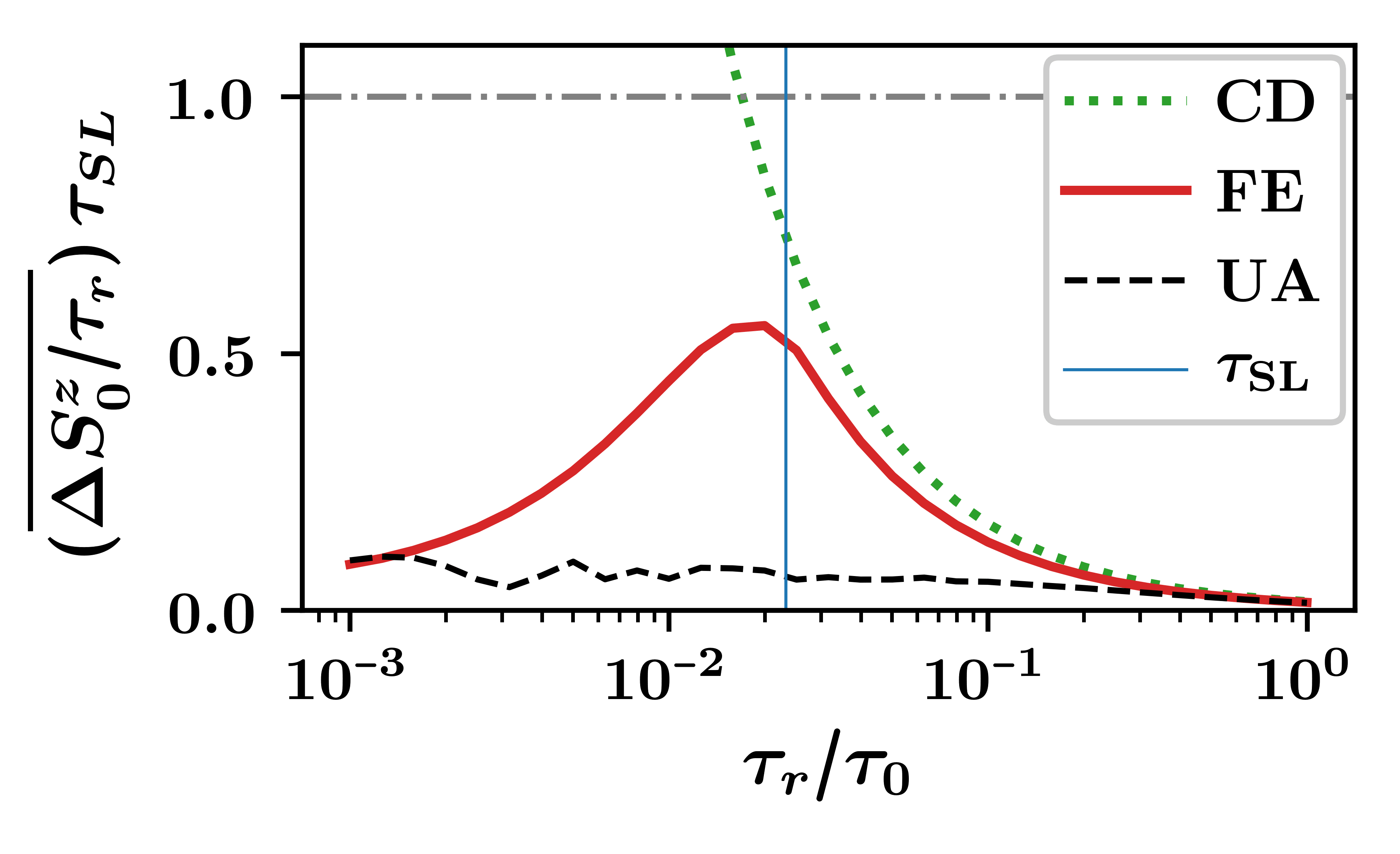}
\caption{ \textbf{Power vs. protocol time.} Disorder-averaged transfer power against the scaled ramp time $\tau_r/\tau_0$ for UA, CD, and FE protocols. The vertical axis is scaled by $(1/\tau_{SL})$ (grey dash-dotted line). The vertical blue solid line marks the disorder-averaged speed limit timescale $\tau_{SL}$. Parameters: $N_s=50$, $L=8$, $\Omega_{B} = 10$, $\Lambda_0 =\lambda_0 = 5$, $\overline{g}=0.1$, $\gamma_{xx} = 0.05$, $\gamma_z = 0.05$, $\tau_0 \approx 1000$, and $\omega=100$.} \label{fig:Power} 
\end{figure}

This speed limit is quantified in Fig.~\ref{fig:Power}, showing the transfer power vs. ramp time for several transfer protocols across resonance. The transfer power is defined as
\begin{equation}
    \frac{\Delta S_0^z}{\tau_r} = \frac{\langle S_0^z(\lambda_f) \rangle - \langle S_0^z(\lambda_i)\rangle}{\tau_r},
\end{equation}
measuring the rate at which polarization is extracted from the qubit per unit ramp time.
In the absence of tunable qubit-bath interactions, the maximum possible polarization transfer is a unit of polarization on the timescale $\tau_{SL}$, shown in the plot as a grey dash-dotted line. 
We find that UA protocols operate far below this rate over the whole range of ramp times. On the other hand, FE protocols significantly enhance power, peaking in the vicinity of the speed limit. 
At protocol durations $\tau > \tau_{SL}$ FE nears the efficiency of CD protocols which, as expected, transfers polarization more effectively than UA and FE per unit time.
At protocol durations $\tau < \tau_{SL}$, the FE power lies significantly below the CD power. 

The presence of this speed limit suggests a broader physical limitation: Any ramped protocol which does not directly tune system-bath couplings or add extra controls cannot transfer polarization at a faster rate than a sudden resonant exchange. 

\section{Conclusion}\label{sec:conclusion}
In this work, we apply the tools of shortcuts to adiabaticity to a class of hyperpolarization protocols. In a single cycle of each such protocol, the qubit is reset along the $-z$ direction by an external pulse, after which the $z$-field of the qubit is swept across a resonance region. Polarization is transferred from the qubit to the spin bath during the sweep.

We introduce local counterdiabatic driving (LCD) protocols that mimic an adiabatic protocol. 
The LCD protocols simultaneously tackle two problems: (i) the small sweep rates necessary for the adiabatic transfer of polarization to the bath, and (ii) the limits on hyperpolarization imposed by dark states. 
The LCD protocols tackle (i) by efficiently suppressing diabatic transitions between bright bands.
They tackle (ii) by depleting dark states in the presence of inhomogeneous bath fields (i.e., when the system is non-integrable). 
In this way, LCD protocols outperform both unassisted protocols and exact counterdiabatic protocols since the former does not suppress transitions between bright bands and the latter suppresses transitions from dark to bright bands.

Using exact numerics and a master equation, we show that the LCD protocols outperform the unassisted ones by various metrics (efficiency, power, and number of cycles).
Additionally, the LCD can be experimentally implemented through a high-frequency Floquet drive on the qubit.
These engineered protocols have a natural quantum speed limit; once the sweep rate exceeds this limit, the LCD protocols cannot be realized through Floquet drives.

The LCD may be used to speed up hyper-polarization in several experimental systems with dipolarly interacting spins. Indeed, Eq.~\eqref{eq:H} models the (rotating-frame) Hamiltonian of a shallow nitrogen-vacancy (NV) defect coupled to surface electronic spins in high-purity diamond \cite{rosskopf_investigation_2014,myers_probing_2014,sushkov_magnetic_2014}, as well as the (rotating-frame) Hamiltonian of a NV defect coupled to bulk C-13 nuclei~\cite{london_polar_2013,scheuer2017robust} or the nuclei of external molecules in solution~\cite{fernandez-acebal_hyper_2017}. A promising avenue for future work is to compare the performance of LCD protocols to sudden protocols that satisfy the Hartmann-Hahn condition~\cite{Hartmann-Hahn} in these systems. Hyperpolarization of powdered diamond using NV-centers~\cite{scheuer2017robust, ajoy2018orientation} is also an important goal for magnetic resonance imaging. It would be interesting to develop LCD protocols that account for the random orientation of the NV center axes (the $z$-direction of the central spin in Eq.~\eqref{eq:H}) in these systems. An additional direction for future work is to use optimal control theory to design possibly more efficient protocols and test the validity of the speed limit. However, such a numerical optimization problem in the many-body setting is expected to be highly complex. In contrast, the LCD approach, while not guaranteed to be optimal, is readily extended to more complex systems with an arbitrary number of spins and arbitrary interactions. 

Theoretically, our work raises questions about the precise interplay between integrability and hyperpolarizability in central spin models. 
The model in Eq.~\eqref{eq:H} with $\gamma_z=0$ is (i) integrable, and (ii) has exact dark states for any choice of $g_j$~\cite{villazon2020integrability}. However, the closely related XXX model with $\gamma_z=0$ and isotropic qubit-bath interactions $\sum_j g_j \vec{S_0} \cdot \vec{S_j}$ is integrable without exhibiting dark states~\cite{gaudin_bethe_2014,dukelsky_colloquium_2004}. The XXX model describes the hyperfine interactions of the electronic spin of a quantum dot with surrounding nuclei~\cite{Hanson:2007aa, schliemann2003electron}. Previous work~\cite{christ_nuclear_2009, imamoglu_optical_2003} suggests that the spin bath can be efficiently polarized in the XXX model despite its integrability. A natural direction for future work is to quantify the general role of integrability in the polarization process.
Another possible direction is to examine the role of interactions in the spin bath. The accompanying diffusive spin transport is expected to aid in the polarization of distant bath spins with negligible $g_j$.

\section*{Acknowledgements.} 
The authors thank E. Boyers, M. Pandey, C.R. Laumann, and A. Sushkov for insightful discussions. 
The authors acknowledge support from the Sloan Foundation through a Sloan Research Fellowship (A.C.), from the Belgian American Educational Foundation (BAEF) through the Francqui Foundation Fellowship (P.W.C.), and from the BU CMT Visitor Program (P.W.C.). Numerics were performed on the BU Shared Computing Cluster with the support of the BU Research Computing Services.
This work was supported by EPSRC Grant No. EP/P034616/1 (P.W.C.), NSF DMR-1813499 (T.V. and A.P.), NSF DMR-1752759 and AFOSR FA9550-20-1-0235 (T.V. and A.C.), and AFOSR FA9550-16- 1-0334 (A.P.).

\appendix

\section{Derivation of central spin Hamiltonian}\label{SI:Model}

Consider a driven qubit-bath spin system in a magnetic field $B$ along the z-direction described by the Hamiltonian:
\begin{equation}\label{eq:Hgen1}
H_{0} = H_Q + H_B + H_D + H_{QB}+H_{BB},
\end{equation}
Here $H_Q = \gamma_{Q} \,B \,S_0^z$ is the Zeeman energy of the central qubit with gyromagnetic ratio $\gamma_Q$, and $H_B = \gamma_{B} \,B \,\sum_{j=1}^{L-1} S_j^z$ is the Zeeman energy of the bath spins with gyromagnetic ratio $\gamma_B$. The driving term is given by
\begin{align}
H_{D} & = 2\,\Omega_{Q}\cos(\omega_Q t)\,S_0^x + 2\cos(\omega_B t)\,\sum_{j=1}^{L-1}\,\Omega_{B,j}\,S_j^x 
\end{align}
where $\Omega_Q$ is a Rabi amplitude on the central spin, $\Omega_{B,j}=\Omega_B + \delta\Omega_{j}$ is a generally inhomogeneous Rabi amplitude on the bath, and $\omega_Q, \omega_B$ are respectively the corresponding driving frequencies. The qubit-bath coupling is given by a dipole interaction term:
\begin{equation}
H_{QB}  =  \sum_i \frac{\gamma_{Q} \gamma_{B}}{r_i^3}\,\bigg[ \vec{S}_0\cdot \vec{S}_i - 3\,(\vec{S}_0\cdot\,\hat{r}_i)\,(\vec{S}_i\cdot\hat{r}_i) \bigg]
\end{equation}
where $\vec{r}_i$ is the vector between the central qubit and the $i^{\mathrm{th}}$ bath spin. The interaction term $H_{BB}$ between bath spin pairs is generally also dipolar. 
In this work, we assume bath-bath interactions are small compared to the qubit-bath couplings, which is realized in experiments with sufficiently low bath spin density or with bath spins that satisfy $\gamma_{B} \ll \gamma_{Q}$.
The Hamiltonian $H_0$ in Eq.~\eqref{eq:Hgen1} describes single qubit systems, such as NV centers in diamond or quantum dots, interacting with an ensemble of spins (e.g. spins on the surface of diamond) and driven by continuous irradiation fields (such as radio waves)~~\cite{hartmann1962nuclear,fernandez-acebal_hyper_2017,rovnyak2008tutorial,rao2019spin,lai2006knight,belthangady2013dressed,cai_simulator_2013}.

In a doubly rotated frame defined by the unitary transformation
\begin{equation}
 U = \exp\bigg[- i\,\bigg(\omega_{Q}\,S_0^z + \omega_{B} \sum_{i=1}^{L-1} S_i^z \bigg) \,t\bigg] ,
\end{equation}
$H_0$ can be simplified by matching the driving frequencies to the Zeeman energies ($\omega_{Q} = \gamma_Q B$, $\omega_{B} = \gamma_B B$), and applying a rotating wave approximation to eliminate rapidly rotating non-secular terms which average to zero on the timescale of the dynamics~\cite{rovnyak2008tutorial}. Relabeling our axes $(x,z) \to (z,-x)$, the dominant time-averaged motion is described by
\begin{equation}\label{eq:Hgen2}
H_{\mathrm{rot}}(t) = \Omega_Q S_0^z + \sum_{j=1}^{L-1} \Omega_{B,j} S_0^z + \sum_{j=1}^{L-1} 2\,g_j\,S_0^x \,S_j^x
\end{equation}
where 
\begin{equation}\label{eq:gj}
g_j \equiv \frac{\gamma_{Q}\gamma_{B}}{2\,r_i^3}[1-3\,\cos^2(\theta_i) ],
\end{equation}
and $\theta_i$ is the angle between $\vec{B}$ and $\vec{r}_i$ in the frame of $H_0$. We note our rotating wave approximation requires $|g_j/B| \ll  \gamma_Q\, ,  \gamma_B\, , |\gamma_Q \pm \gamma_B|\, $, which is readily satisfied in NV centers or quantum dot experiments~\cite{belthangady2013dressed,gullans2013preparation}. 

The interaction term 
\begin{equation}
S_0^x S_j^x = \frac{1}{4}\big(S_0^+ S_j^- + S_0^- S_j^+ + S_0^+ S_j^+ + S_0^- S_j^-\big)  
\end{equation} 
describes zero quantum (flip-flop) transitions in its first two terms, and double quantum (flip-flip/flop-flop) transitions  interactions in its last two terms. Zero quantum transitions dominate when $g_j \ll \Omega_Q + \Omega_{B,j}$~\cite{rovnyak2008tutorial}, yielding the Hamiltonian presented in the main text:
\begin{equation}\label{eq:A1}
H(t) = \Omega_Q \,S_0^z + \sum_{j=1}^{L-1} \Omega_{B,j}\, S_j^z + \frac{1}{2}\sum_{j=1}^{L-1} g_j\, \big(S_0^{+} S_j^{-} + S_0^{-} S_j^{+} \big).
\end{equation}

\section{Distribution of energy gaps in the homogeneous limit}\label{SI:HomoGaps}

In this section, we compute the approximate distribution of bright pair gaps $\Delta_{\alpha}$ (Eq.\eqref{eq:AnalyticGaps}) in a system with a homogeneous bath field ($\gamma_z=0$).

In the homogeneous limit ($\gamma_{xx}=0$), the central spin model reduces to a two-body Hamiltonian
\begin{equation}
H = \lambda S_0^z + \frac{\overline{g}}{2} \left(S_0^+ S^- + S_0^- S^+\right),
\end{equation}
with $S^{\pm} = \sum_{j=0}^{L-1} S_j^{\pm}$. The spectrum of this Hamiltonian can be obtained in the collective bath spin basis, as the Hamiltonian only couples the states:  $\ket{\uparrow} \otimes \ket{s,m}, \ket{\downarrow} \otimes \ket{s,m+1}$. Here $s$ is the total spin quantum number of the bath, and $m$ is the total z-projection of the bath state, leading to $M = m + 1/2$. We take $-s < m < s$, since the states $\ket{\uparrow} \otimes \ket{s,s}$ and $\ket{\downarrow} \otimes \ket{s,-s}$ are dark eigenstates of the Hamiltonian.

The energy in each two-dimensional subspace is fully determined by the quantum numbers $s$ and $m$, leading to energies $\pm \Delta_{s,m}/2$ at resonance given by
\begin{equation}
\Delta_{s,m} =\overline{g} \sqrt{(s-m)(s+m+1)}.
\end{equation}
The number of gaps equal to $\Delta_{s,m}$ is fully determined by the number of ways the $(L-1)$ spin-$1/2$ bath spins can be coupled to a collective spin $s$ with spin projection $m$. Furthermore, since $m < s$ and $m$ is fixed by specifying $M$, this also leads to a minimal gap within each polarization sector $M$, given by $\Delta_{m} = \overline{g}\sqrt{2M+1}$, obtained by setting $s=m+1=M+1/2$. Increasing $s$ resulting in an increasing $\Delta_{s,m}$, whereas smaller values of $s$ are not allowed within this polarization sector.

Given $(L-1)$ bath spins, the number of spin-$s$ representations is given by Catalan's triangle as
\begin{align}
N_s(L) &= C\left((L-1)/2+s,(L-1)/2-s\right) \\
&= \frac{(L-1)!(2s+1)}{(L/2-1/2-s)!(L/2+1/2+s)!},
\end{align}
which can be approximated for large $L$ as
\begin{equation}
N_s(L) \approx \frac{e^{L f_s(1/2-s/L)}}{\sqrt{2 \pi}} \frac{2s+1}{L/2+s+1} \sqrt{\frac{L}{(L/2-s)(L/2+s)}},
\end{equation}
with $f_s(p) = -p \ln(p) - (1-p) \ln(1-p)$. The total magnetization $M$ fixes $m$ such that the energy gap only depends on $s$, and we can introduce a gap density as
\begin{equation}
n(\Delta) = N_{s(\Delta)}(L) \frac{ds}{d\Delta},
\end{equation}
with $s(\Delta) = \sqrt{(\Delta/\overline{g})^2 +M^2}-1/2$ and
\begin{equation}
\frac{ds}{d\Delta} = \frac{\Delta/\overline{g}^2}{\sqrt{(\Delta/\overline{g})^2 +M^2}}.
\end{equation}
As mentioned before, every fixed magnetization sector has a minimal gap, which we write as $\Delta_m \equiv \overline{g}\sqrt{2M+1}$, such that $n(\Delta < \Delta_m) = 0$. Due to the presence of the exponential term, in the limit of large $L$ all integrals over $n(\Delta)d\Delta$ will be dominated by the boundary terms where $\Delta \approx \Delta_m$. Approximating the exponential factor at $s/L = |M|/L \equiv \tilde{m}$ for $\Delta_m < \Delta \ll \overline{g}M$, we find
\begin{equation}
n(\Delta) \approx K(\tilde{m}) \frac{e^{L f_s(1/2-\tilde{m})}}{\sqrt{2\pi L}} \frac{\Delta}{\Delta_m^2}\left(\frac{1-2\tilde{m}}{1+2\tilde{m}}\right)^{\frac{\Delta^2}{\Delta_m^2}},
\end{equation}
with 
\begin{equation}
K(\tilde{m}) = \frac{4\tilde{m}}{1/2+\tilde{m}}\sqrt{\frac{1}{1/4-\tilde{m}^2}}.
\end{equation}

\section{Diabatic transitions in the Landau-Zener problem}\label{SI:LZ}

The Landau-Zener (LZ) problem, described in Eq.~\eqref{eq:brightHam} of the main text, consists of a two-level system with gap  $\Delta_{LZ} = \sqrt{\lambda^2 +\Delta^2}$,
where $\lambda$ is a control field and $\Delta$ is the minimum gap~\cite{AP1,Shevchenko}.

When the control field is varied at a speed $\dot{\lambda} \sim \lambda_0/\tau_r$, we can estimate the speed scale below which the system remains adiabatic. Adiabaticity occurs when the rate of change of the gap $\Delta$ is smaller than the dynamical timescale over the whole range of the control field $\lambda$. In particular, this condition holds if it is satisfied  near resonance ($|\lambda|\sim\Delta$) where the gap is smallest. 
\begin{align}
\bigg(\frac{\dot{\Delta}_{LZ}}{\Delta_{LZ}} \ll \Delta_{LZ} \,\bigg)\,\bigg|_{\lambda=\Delta}\implies \dot{\lambda} \ll \Delta^2.
\end{align}
Therefore we satisfy the adiabatic condition when the ramp timescale $\tau_r \gg \lambda_0/\Delta^2$. In faster ramps moreover, the scale $\tau_0 \sim \lambda_0/\Delta^2$ sets the scale for the onset of diabatic transitions.

As discussed in the main text for our central spin model, the LZ problem directly captures the interactions between bright bands. Nevertheless, the LZ problem can also help understand transitions between dark and bright bands, as we discuss next. 

In system with $\gamma_z=0$, the gap $\Delta_{DB}$ between a dark state with energy $E_D$ and a neighboring bright state with energy $E_B$ is given by  
\begin{align}
\Delta_{DB} = |E_B - E_D| = \frac{1}{2}\sqrt{\lambda^2+\Delta_{\mathrm{min}}^2} \pm  \frac{1}{2}\lambda,
\end{align}
since $E_D = \pm\lambda/2$ and $E_B = \pm \frac{1}{2}\sqrt{\lambda^2+\Delta_{\mathrm{min}}^2}$ and where $\Delta_{\mathrm{min}} \equiv \mathrm{min}_{\alpha} \Delta_{\alpha}$ is the minimum bright-bright gap at resonance.
At resonance, the gap is $\Delta_{DB} = \Delta_{\mathrm{min}}/2$, comparable to the minimum bright-bright gap. In contrast with bright-bright gaps, dark-bright gaps are smallest furthest away from resonance $\lambda = \pm\lambda_0$:
\begin{align}\label{eq:DeltaDBmin}
\mathrm{min}\,\Delta_{DB} \approx \frac{1}{4}\bigg(\frac{\Delta_{\mathrm{min}}}{\lambda_0}\bigg)^2 \lambda_0.
\end{align}
In systems with $\gamma_z=0$, the presence of a small bright-dark gap does not imply fast ramps yield diabatic transitions because the driving operator $S_0^z$ does not couple bright and dark states. 

When z-disorder $\gamma_z>0$ is introduced in the bath field, dark and bright bands mix and $\gamma_z$ sets an energy window for perturbed dark/bright energies.
Then the perturbed dark and bright bands will begin to overlap at a critical z-disorder $\gamma_z^c$ given by 
\begin{equation}
\gamma_z
^c \sim \mathrm{min}\,\Delta_{DB} \sim \Delta_{\mathrm{min}}^2/\lambda_0.
\end{equation}
For $\gamma_z\gtrsim\gamma_z^c$, the gap between dark and bright bands is effectively closed away from resonance. Then dark-bright transitions depend on the many-body level spacing $\Delta E\ll \mathrm{min}\,\Delta_{DB}$.
Therefore, any finite-speed ramps will yield dark-bright diabatic transitions provided 
\begin{equation}
    \tau_r \lesssim \lambda_0/(\Delta E)^2.
\end{equation}
On the other hand, for small but finite z-disorder regime $0 <\gamma_z \ll \gamma_z^c$, Eq.~\eqref{eq:DeltaDBmin} accurately approximates the dark-bright minimum gap. The ramp timescale for the onset of diabatic dark-bright transitions is then given by:
\begin{align}
    \tau_r \sim \lambda_0/(\mathrm{min}\,\Delta_{DB})^2 \sim \lambda_0^3/\Delta_{\mathrm{min}}^4.
\end{align}

\section{LCD with a gapless model}\label{SI:Gapless} 

\begin{figure}[t]
\centering
   \includegraphics[width=1.0\columnwidth]{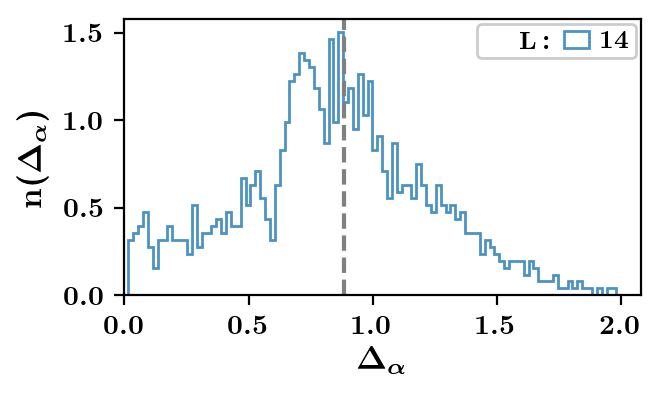}
\caption{  \textbf{Histogram of resonant gaps at large xx-disorder strength.} Plot shows histogram of resonant gaps for $\gamma_{xx}=0.5$ at $L=14$ and $M=-1$. Vertical dashed grey lines shows reference scale $\Delta_{\mathrm{typ}}$.
Parameters: $N_s=1$, $\Omega_{B}=10$, $M=-1$, $\lambda= 0$, $\overline{g}=0.1$,$\gamma_z = 0.00$, $L=14$.  } \label{fig:GaplessDistribution}
\end{figure}

Our main work has focused on the weak disorder limit $\gamma_{xx},\gamma_{z} < \overline{g}$, in which there is a clear non-zero gap between the bright bands at numerically accessible system sizes. Although there is weak trend suggesting that the gap may close in the thermodynamic limit, it is not clear from the numerics whether and how this gap will close. In this section, we show that the LCD protocols are still efficient for polarization transfer and depopulating dark states in a gapless system. 

To probe a gapless system, we consider the strong xx-disorder limit $\gamma_{xx}\gg \overline{g}$ of the Hamiltonian model in equation~\eqref{eq:H}. Then bright pair gaps have the the gap distribution shown in Fig.~\ref{fig:GaplessDistribution}. Although many bright pairs show exponentially small gaps, the bulk of the distribution lies on the energy scale $\Delta_{\mathrm{typ}} = \sqrt{\sum_j g_j^2}$, as shown by the dashed grey vertical line.

\begin{figure}[t]
\centering
   \includegraphics[width=1.0\columnwidth]{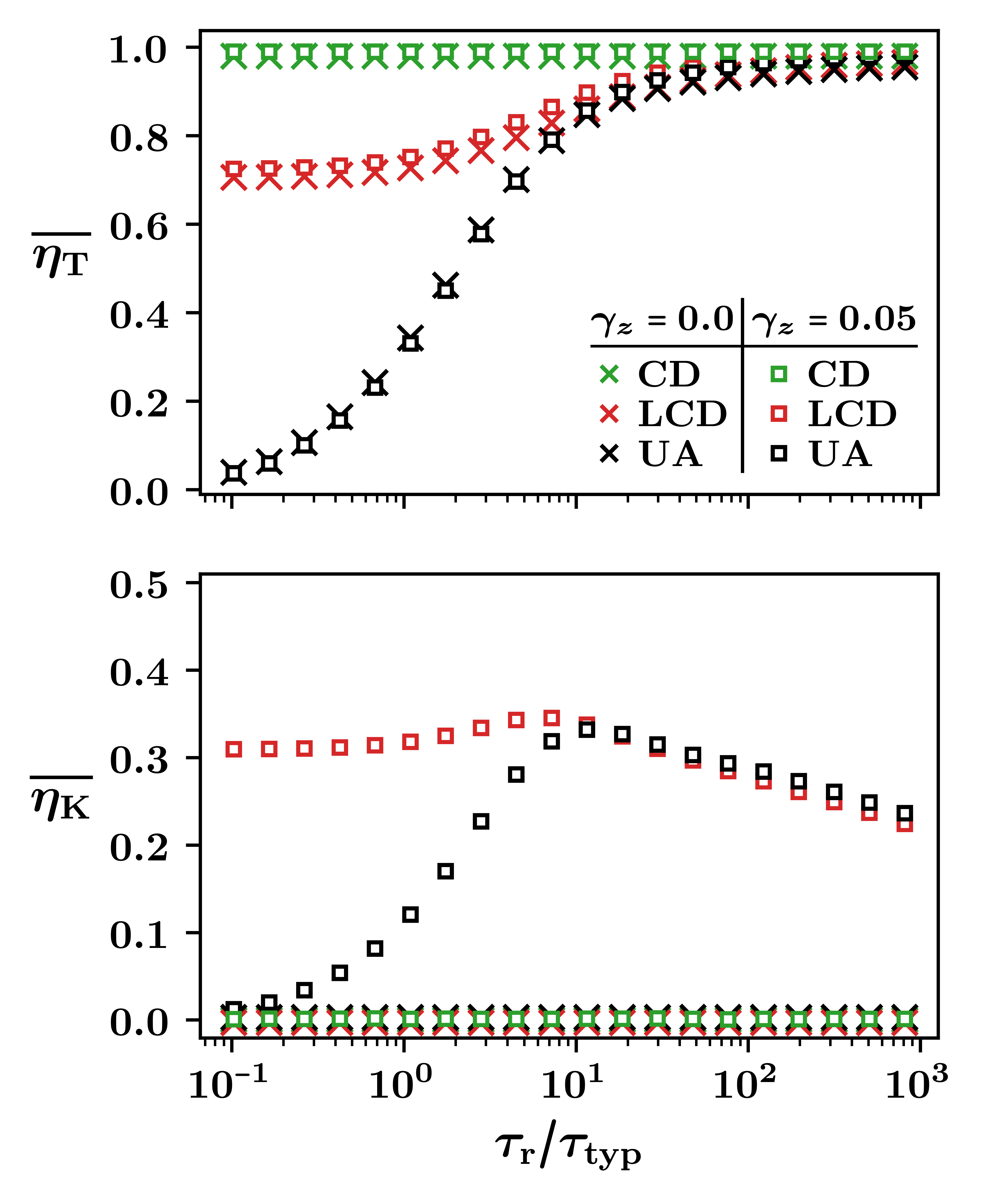}
\caption{ \textbf{Efficiency vs. ramp time at large xx-disorder.} Disorder-averaged transfer efficiency (top) and kick efficiency (bottom) of UA, CD, and LCD protocols in systems with $\gamma_z=0.00$ (crosses) and $\gamma_z=0.05$ (squares).% Dashed grey line indicates the theoretical prediction based on Eq~\eqref{eq:EffErr}.
 Parameters: $N_s = 100$, $L=10$, $M = -1$, $\Omega_{B} = 10$, $\lambda_0 = 5$, $\overline{g}=0.1$, $\gamma_{xx} = 0.5$, and $\tau_{\mathrm{typ}} \approx 10$. } \label{fig:Eff-large-xx-disorder}
\end{figure}

Fig.~\ref{fig:Eff-large-xx-disorder} shows the transfer and kick efficiency for CD, LCD, and UA protocols as a function of ramp time $\tau_r$ at large xx-disorder $\gamma_{xx} = 5$, $\overline{g} =0.5$. For concreteness, we focus on the magnetization sector $M=-1$ with large Hilbert dimension to bring out the exponential closing of the smallest gaps. Since the minimum bright pair gap is exponentially small\footnote{For comparison with the main text, at weak xx-disorder, $\Delta_{\mathrm{min}} \sim \overline{g}$ in this magnetization sector at accessible system sizes.} in $L$, making $\tau_0\sim \Delta_{\mathrm{min}}
^{-2}$ exponentially large, we rescale the horizontal axis instead by $\tau_{\mathrm{typ}} \approx 2\lambda_0/\sum_j g_j^2$, which sets the scale for the onset of diabatic transitions between a typical bright pair in the bulk of the spectrum.
The behavior of all protocols is qualitatively similar to the weak disorder limit. In particular, our LCD protocol maintains a relative high transfer efficiency and non-zero kick efficiency at fast ramp speeds. This is to be expected since LCD suppresses transitions by targeting a single bright pair gap $\Delta_{\mathrm{typ}}$, which is the gap scale for the bulk of bright pairs.

\section{Breaking integrability with z-disorder}\label{SI:Chaos}

\begin{figure}[t]
\centering
   \includegraphics[width=1.0\columnwidth]{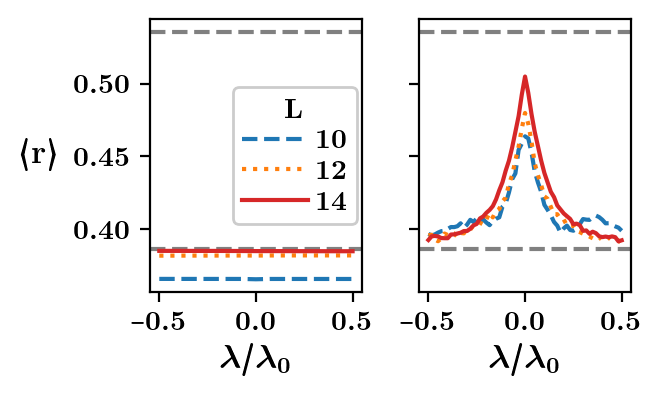}
\caption{ \textbf{Average level spacing ratio vs. qubit field detuning.} The average ratio $\langle r \rangle$ is shown for several system sizes $L$. Left panel (homogeneous bath field) shows a trend toward Poisson statistics with increasing system size. Right panel (inhomogeneous bath field) reveals a trend toward GOE statistics around resonance. Grey dashed lines indicate the Poisson and GOE values.
Parameters: $N_s=100$, $M=-1$, $\Omega_{B} = 10$, $\lambda_0= 5$, $\overline{g}=0.1$, $\gamma_{xx} = 0.05$, $\gamma_z = 0.0$ (left), $\gamma_z=0.05$ (right). } \label{fig:LevelStats}
\end{figure}

Energy level statistics are a widely used diagnostic for ergodicity and chaos~\cite{poilblanc1993poisson,casati1985energy,atas2013distribution}. 
The average level spacing ratio $\langle r \rangle$ is obtained by averaging over an ordered distribution of energies $\{E_n\}$ the ratio $r_n = \min(s_n,s_{n-1})/\max(s_n,s_{n-1})$, where $s_n = E_{n+1}-E_{n}$~\cite{oganesyan2007localization, atas2013distribution}. In integrable systems satisfying a Poisson distribution of energy spacings, $\langle r \rangle \approx 0.3863$, while ergodic systems are expected to satisfy a Wigner-Dyson distribution in accordance to a grand orthogonal ensemble with $\langle r \rangle \approx 0.5307$. 

Fig.~\ref{fig:LevelStats} shows  $\langle r \rangle$  for a range of detunings $\lambda$ around resonance and multiple system sizes. We average the level spacing ratio over the middle two quartiles of the spectrum in the sector $M=-1$ and further average over $N_s=100$ disorder realizations. The left panel shows a system with $\gamma_z=0$, where the ratio $\langle r \rangle$ tends to the Poisson value with increasing system size over the whole range of detunings. In contrast, the right panel shows a system with $\gamma_z=0.05$, where $\langle r \rangle$ tends to the Wigner-Dyson value with increasing system size around resonance $\lambda\approx 0$. This behavior suggests that $\gamma_z>0$ breaks the integrability of model and establishes chaos and ergodicity.

\section{Gauge potential in the absence of z-disorder}\label{SI:CD}

In the presence of a homogeneous global bath field $\Omega_B$, the central spin Hamiltonian $H$ is integrable and varying $\lambda(t)$ constitutes an integrable perturbation~\cite{villazon2020integrability}. Then the closed form of the gauge potential Eq.~\eqref{eq:Aclosed} is obtained from Eq.~\eqref{expansion} as follows. Let $\Delta H \equiv H - \Omega_B M$. Then it can be checked that
\begin{align}
&[ H , S_0^z ] = [ \Delta H , S_0^z ] = i\,\sum_j g_j (S_0^x \,S_j^y - S_0^y\, S_j^x), \label{com1}\\ 
&\Delta H [\Delta H, S_0^z ] = - [\Delta H, S_0^z ] \,\Delta H. \label{com2}
\end{align}
The exact gauge potential satisfies \cite{Kolodrubetz}
\begin{equation}
[\partial_{\lambda}H + i [\mathcal{A}_{\lambda},H],H] = 0,
\end{equation}
which can be solved for $\partial_{\lambda}H  = S_0^z$ using Eq.~\eqref{com2} by
\begin{equation}
\mathcal{A}_{\lambda} = -\frac{i}{4}\,(\Delta H)^{-2}\,[\Delta H, S_0^z].
\end{equation}
The pseudo-inverse only acts on the bright states, where we can expand it as
\begin{equation}
(\Delta H)^{-2} = 4 \sum_{\alpha}\frac{ \mathbbm{1}_{\alpha}}{\lambda^2 + \Delta_{\alpha^2}}.
\end{equation}
This reduces to a constant in each bright-state subspace, and returns the total gauge potential as
\begin{equation}
\mathcal{A}_{\lambda} = -\sum_{\alpha}\frac{ \Delta_{\alpha}}{\lambda^2 + \Delta_{\alpha^2}} \tilde{S}_{\alpha}^y.
\end{equation}
An alternative way of motivating this solution is by noting Eq.~\eqref{com2} implies that
\begin{align}
[ \underbrace{H, [H, \dots [H}_{k}, [H,S_0^z] ] &= (2 \Delta H)^k \,[\Delta H, S_0^z ]= 2\, \Delta H [\Delta H, S_0^z ]\nonumber\label{com3}
\end{align}
Since all nested commutators are proportional to $[\Delta H, S_0^z]$, the commutator expansion for the gauge potential implies that the gauge potential itself needs to be proportional to this commutator up to a $\Delta H$-dependent prefactor \cite{claeys2019floquet}.

\section{Local variational  approximations of the gauge potential}\label{SI:LCD}

\begin{figure}[hb]
\centering
   \includegraphics[width=1.0\columnwidth]{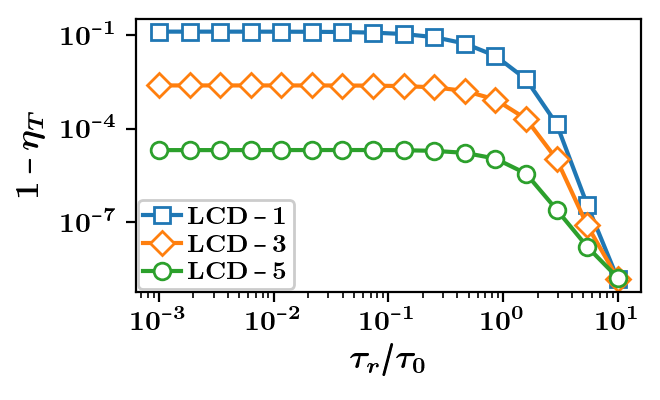}
\caption{ \textbf{Transfer error vs. ramp time for higher order LCD.} 
Curves are shown for LCD in the leading three orders labeled by the number of occurrences of the $H$ in $[H,[H,\dots,[H,\partial_{\lambda}H]]]$. Higher order curves progressively approach the exact CD result $1-\eta_T = 0$.
Parameters: $L=8$, $N_s=1$, $M=-1$, $\Omega_{B} = 10$, $\lambda_0= 5$, $\overline{g}=0.1$, $\gamma_{xx} = 0.05$, $\gamma_z = 0.0$, $\tau_0=1000$. } \label{fig:ConvergenceLCD} 
\end{figure}

The gauge potential can approximated by truncating Eq.~\eqref{expansion} to a desired order $q$~\cite{claeys2019floquet}. The approximation can be improved by setting new coefficients $\alpha_j$ which variationally minimize the following action~\cite{Sels2017}:
\begin{equation}
\mathcal{S}(\mathcal{A}_{\lambda}) = \mathrm{Tr}[G^2]; \quad\quad G\equiv \partial_{\lambda}H + i[\mathcal{A}_{\lambda},H].
\end{equation}
First we express $\mathcal{A}_{\lambda} = \sum_j \alpha_j A_j$, where $\{A_j : j=1,\dots,q\}$ is the operator basis of order $q$ with $$A_j = \underbrace{[H,\dots,[H}_{2j-1},\partial_{\lambda}H]]].$$
Differentiating $\mathcal{S}(\mathcal{A}_{\lambda})$ with respect to $\alpha_k$, we obtain:
\begin{equation}\label{eq:alphaMat}
    \sum_j \alpha_j\, \mathrm{Tr}\big[\{[H,A_j],[H,A_k] \}\big] = \mathrm{Tr}\big[\{\partial_{\lambda}H,[A_k,H] \}\big]
\end{equation}
Thus solving for the variational coefficients $\alpha_j$ is equivalent to solving a matrix equation. In the simplest case (order $q=1$) we can solve the resulting equation directly:
\begin{equation}
    \alpha_1 = \frac{\mathrm{Tr}\big[\{\partial_{\lambda}H,[[H,\partial_{\lambda}H],H] \}\big]}{2\,\mathrm{Tr}\big[[H,[H,\partial_{\lambda}H]]^2 \big]}.
\end{equation}
For our purposes, $\partial_{\lambda} H = S_0^z$, and this expression can be evaluated analytically to yield
\begin{equation}
\alpha_1 =-\frac{1}{\lambda^2+\sum_j g_j^2},
\end{equation}
in accordance with Eq.~\eqref{ALCD} presented in the main text.

Higher order approximations can be obtained by solving Eq.~\eqref{eq:alphaMat} analytically or numerically. Fig.~\ref{fig:ConvergenceLCD} showcases the progressive convergence of local CD approximations to the exact CD as order $q$ is increased. The plot shows the transfer efficiency for a numerical simulation of a sweep across resonance over several orders of magnitude in ramp time. Although higher order approximations mimic CD more closely, they come at the cost of increased complexity which may not be experimentally feasible.

\section{High Frequency Floquet Hamiltonian}\label{SI:FE}

We detail the stroboscopic equivalence between the Floquet Hamiltonian $H_F$ and the LCD Hamiltonian in Eqs.~\eqref{eq:HFE_2LS} and \eqref{FE} to leading order in the limit of high frequency $\omega$.

Considering the two-level Hamiltonian \eqref{eq:HFE_2LS}, we first go to a rotating frame to cancel the rapidly oscillating term $\propto \beta(t) \omega sin(\omega t)$ that will be dominant in the limit $\omega \to \infty$. Given a unitarity transformation $U$, the effective Hamiltonian in the moving frame is given by
\begin{equation}
\bar{H}_{\mathrm{FE}} = U^{\dagger} H_{\mathrm{FE}} U - i U^{\dagger} \partial_t U.
\end{equation}
We now take
\begin{equation}
U = \exp\bigg( - i\, \theta(t)\, S^z  \bigg),
\end{equation}
generating a rotation about the z-axis by an angle $\theta(t) =  \beta(t)\,(1-\cos(\omega\,t))$. The resulting Hamiltonian in this rotating frame reads:
\begin{align}
\bar{H}_{\mathrm{FE}} = \gamma(t) S^z + \Delta \left[\cos(\theta(t)) S^x - \sin(\theta(t))S^y\right]. 
\end{align}
In the high-frequency limit, $\omega^{-1}$ is the fastest timescale, leading to a time-scale separation such that all parameters in the Hamiltonian are effectively constant over multiple driving periods of $\cos(\omega\,t)$. Moreover, we can perform a Magnus expansion of $\bar{H}_{FE}$ in powers of $\omega^{-1}$ to approximate the Floquet Hamiltonian $H_F$~\cite{Bukov2015}. To leading order, $H_F$ is found by averaging $\bar{H}_{FE}$ over a period $T = 2\pi/\omega$:
\begin{align}\nonumber
H_{\mathrm{F}} = \bar{H}_{\mathrm{FE}} [\cos(\theta) \to \langle\cos(\theta)\rangle; \sin(\theta)\to \langle\sin(\theta)\rangle] + \mathcal{O}\bigg(\frac{1}{\omega}\bigg)
\end{align}
where 
\begin{align}
\langle\cos(\theta)\rangle = \frac{1}{T} \int_0^T dt \cos(\theta(t)) \approx J_0(\beta) \cos(\beta),
\end{align}
and
\begin{align}
\langle\sin(\theta)\rangle = \frac{1}{T} \int_0^T dt \sin(\theta(t)) \approx J_0(\beta) \sin(\beta),
\end{align}
where $J_0$ is a Bessel function of the first kind, and any time-dependence of $\beta$ has been neglected. This returns the Floquet Hamiltonian proposed in the main text, satisfying $H_{\mathrm{F}} = G(\beta) H_{LCD}$ with $G(\beta) = J_0(\beta) \cos(\beta)$.

The same derivation holds for the central spin model, with the unitary transformation given by $U = \exp ( - i\, \theta(t)\, S_0^z )$. This returns
\begin{align} \nonumber
\bar{H}_{\mathrm{FE}} &= G(t)\,\lambda(s(t))\,S_0^z + \sum_j \delta\Omega_j S_j^z \\ \nonumber
&+ \sum_j  g_j\,[ \cos(\theta(t))\, S_0^x - \sin(\theta(t))\, S_0^y] \,S_j^{x}  \\ 
& + \sum_j \,g_j\,[  (\sin(\theta(t)) S_0^x + \cos(\theta(t)) S_0^{y}]\,S_j^{y}.
\end{align}
Regrouping terms and using Eq.~\eqref{Comm1} of the main text, we obtain the presented Floquet Hamiltonian as the period-averagd $\bar{H}_{\mathrm{FE}}$.

We conclude this section with a couple of remarks: (i) The lab and rotating frames periodically coincide such that the state of the system under dynamics generated by $H_{\mathrm{FE}}(t)$ is stroboscopically identical in both frames. (ii) The evolution under $H_{\mathrm{FE}}(t)$ is further stroboscopically equivalent to the evolution under $H_{\mathrm{F}}(t)$ by design, where the remaining time dependence in $H_{\mathrm{F}}$ is through the slow variation of $\beta(t)$ and $\gamma(t)$. This implies by extension that dynamics under the LCD captured by $H_{\mathrm{F}}(s)$ is stroboscopically equivalent to $H_{\mathrm{FE}}(t)$ to leading order at high driving frequencies. (iii) Requiring that $\dot{\lambda} = \ddot{\lambda} = 0$ at the ramp endpoints further guarantees the equivalence and stability of FE and LCD over a full sweep across resonance. \\

\section{Speed Limit}\label{SI:SpeedLimit}

The mapping between $H_{\mathrm{FE}}$ and $H_{\mathrm{LCD}}$ requires a time-rescaling transformation $dt\to ds = G(t) dt$. Interestingly, there exists a finite lab timescale $\tau_{SL}$ at which the corresponding rescaled time vanishes. Although FE ramps over shorter timescales are possible, they no longer map to LCD ramps.

To derive the speed-limit timescale consider a lab frame ramp with timescale $\tau$ and let $\tau_S $ be the corresponding ramp in the rescaled frame.
Then
\begin{equation}\label{taus}
\tau_S = \int_0^{\tau} G(t) \,dt , \quad\quad \tau = \int_0^{\tau_S} [G(t(s))]^{-1} ds.
\end{equation}
Since $G(t) < 1$, then $\tau_S <\tau$. The smallest lab timescale for which this mapping holds is given by:
\begin{equation}
\tau_{SL} = \lim_{\tau_S\to0}\int_0^{\tau_S} [G(t(s))]^{-1} ds.
\end{equation}
Re-scaling the integral by $s'= s/\tau_S$ and expanding the $J_0(\beta)\,\cos(\beta)$ term to leading order for $\beta \approx \pi/2$, we obtain:
\begin{equation}
    \tau_{SL} = -\lim_{\tau_S\to0} C\int_{0}^{1} \tau_S \,\dot{\lambda}(\tau_S \,s') \,\alpha_{1}(\tau_S\,s') \,ds' + \mathcal{O}(\tau_S)
\end{equation}
where
\begin{equation}
   C= \frac{J_0(\pi/4)}{\sqrt{2}\,J_0(\pi/2)}\approx 1.28.
\end{equation}
For concreteness, we evaluate this limit directly for a linear ramp $\dot{\lambda} = 2\lambda_0/\tau_S$. Using Eq.~\eqref{ALCD}, we have
\begin{equation}
    \tau_{SL} = 2\,\lambda_0\,C\int_{0}^{1} \frac{1}{(2\,\lambda_0\,s' - \lambda_0)^2 + \sum_j g_j^2}  \,ds'.
\end{equation}
Setting $\Delta_{\mathrm{typ}} = \sqrt{\sum_j g_j^2} = \sqrt{(L-1)}\overline{g}$, we obtain
\begin{equation}
    \tau_{SL} \approx  1.28\, \frac{\arctan[2\lambda_0/\Delta_{\mathrm{typ}}]}{\Delta_{\mathrm{typ}}} \sim \frac{1}{\sqrt{L-1}\,\overline{g}},
\end{equation}
where the last scaling is obtained for ramps which begin/end far from resonance $\lambda_0\gg \Delta$, and using $g_{\mathrm{rms}}\sim \overline{g}$ for $\gamma_{xx} \lesssim \overline{g}$.

%\section{Reset Rates}\label{SI:ResetRates}

%\bibliographystyle{plain}
\bibliography{references}
\end{document}